\documentclass[sigconf]{acmart}
\AtBeginDocument{%
  }

\copyrightyear{2026}
\acmYear{2026}
\acmDOI{XXXXX}
\acmConference[]{}{}{}
\acmISBN{}

\usepackage{array}           
\usepackage{longtable}       
\usepackage{boldline}        
\usepackage{hhline}          

\usepackage{colortbl}
\usepackage{fontawesome5}
\usepackage{makecell}
\usepackage{multirow}
\usepackage{subcaption}
\usepackage{url}

\newcommand{\ignoreblock}[1]{}

\newcolumntype{L}[1]   {>{\raggedright\let\newline\\\arraybackslash\hspace{0pt}}p{#1}}
\newcolumntype{C}[1]   {>{\centering\let\newline\\\arraybackslash\hspace{0pt}}p{#1}}
\newcolumntype{R}[1]   {>{\raggedleft\let\newline\\\arraybackslash\hspace{0pt}}p{#1}}

\usepackage{pifont}  

\begin{document}

\title[Classifying by Proxy: Ensemble of Proxy Tasks for CSAI classification]{Classifying by Proxy: Explainable and Reproducible Ensemble of Proxy Tasks for Child Sexual Abuse Imagery Classification}

\author{Clara Ernesto$^{1}$, Carlos Caetano$^{2}$, Sandra Avila$^{2}$, João Macedo$^{3}$, Camila Laranjeira$^{4}$, Leo S. F. Ribeiro$^{1}$\vspace{0.1cm}}

\affiliation{
            \institution{$^{1}$Instituto de Ciências Matemáticas e de Computação (ICMC), Universidade Estadual de São Paulo (USP), Brazil}}

\affiliation{
            \institution{$^{2}$Instituto de Computação (IC), Universidade Estadual de Campinas (UNICAMP), Brazil}}

\affiliation{
            \institution{$^{3}$Departamento de Ciência da Computação, Universidade Federal de Minas Gerais (UFMG),  Brazil}}

\affiliation{
            \institution{$^{4}$Instituto Federal de Educação, Ciência e Tecnologia de Minas Gerais (IFMG), Brazil\vspace{0.5cm}}}

\renewcommand{\shortauthors}{Clara Ernesto et al.}

\begin{abstract}
Child Sexual Abuse Imagery (CSAI) classification systems are needed solutions for lessening the psychological impacts often felt by law enforcement agents responsible for evaluating these materials and for efficient removal of these materials from the web. However, due to the nature of the task, researching and developing such systems is not a trivial endeavor. The images are highly sensitive, and the related datasets are under restrictive access regimes, which means most studies in the area are not reproducible or distributable and are therefore hard to compare and validate. More concerning still, most models for this task today lack an aspect often desired by law enforcement agents: explainability. In this paper, we apply an ensemble of Proxy Tasks --- tasks that correlate to CSAI classification --- yielding improvements in reproducibility, explainability, and security for distribution. This concept is applied for the first time to real CSAI, with a novel selection of relevant Proxy Tasks (selected from the CSAI literature) and training adaptations to the original framework. Our final model achieves competitive results, yielding 91.9\% balanced accuracy on the RCPD dataset with the best Proxy Task combination. We furthermore contrast these results with the best-in-class representation learning model, DINO, and show that our ensemble improves accuracy and provides explanations for its classification results, a feature that a single deep learning model can seldom provide. 
\end{abstract}


\maketitle

\section{Introduction} 

United Nations Children's Fund (UNICEF) estimates that around $20\%$ of girls and $14\%$ of boys have been subjected to sexual violence worldwide~\cite{unicef2024numbers}.
The report classifies the types of abuse into physical contact, e.g., rape, and non-physical, including sexual harassment, unwanted exposure of body parts, and non-consensual image taking. 
Sexual abuse may bring lasting consequences, such as depression, post-traumatic stress, and substance abuse~\cite{HORNOR2010358}. These impacts are intensified by the production and sharing of Child~Sexual~Abuse~Imagery~(CSAI), exposing and perpetuating the violence suffered by victims.

Law enforcement agents (LEA) and child protection institutions are tasked with reviewing massive volumes of content seized from suspects or received from reports, often incurring in deep psychic strain~\cite{sanchez2019practitioner}. Automating CSAI classification would therefore both lessen their burden and combat the heavy trauma inflicted upon the victims.

In response to this need, tools for CSAI classification have modernized and evolved. However, many steps in model development face ethical, technical and legal difficulties, mainly due to the data's sensitive nature. For instance, access to CSAI datasets is, as it should be, restricted by law in many countries; Partnerships with LEA are then essential for proper handling of these materials, as these specialized personnel are both trained and permitted to do so. We are fortunate to have access to such partnerships, and this study is further enriched by their contributions and input as stakeholders and users of these systems. In addition to access restrictions, models trained directly on CSAI are vulnerable to model inversion~\cite{zhang2020secretrevealergenerativemodelinversion}, as sensitive information from CSAI is represented in its parameters. With this in mind, models trained directly on the target domain should not be publicly available. These restrictions mean that studies can seldom be scrutinized by other researchers, with both model and data incurring risks if distributed openly.

Having these constraints in mind, Proxy Tasks~\cite{Coelho2025MinimizingRisks} were considered a viable solution for building systems that can be developed, tested, and distributed without exposing sensitive information. \textit{Proxy Tasks} are defined as sub-tasks related to the target task that are openly accessible and usable, given the existing restrictions in the target context. For CSAI classification, commonly leveraged sub-tasks include age estimation and pornography classification~\cite{laranjeira2022seeinglookinganalysispipeline}, and law enforcement restrictions include limited access to specialized hardware~\cite{sanchez2019practitioner} and the requirement to execute the entire pipeline locally, thus favoring lightweight models.

Our work builds upon and addresses the limitations of our work-in-progress proposal~\cite{EISPErnesto2025}. We presented a framework called Ensemble for Inference in Sensitive data using Proxy Tasks (EISP), with the goal of leveraging Proxy Task features for inference on target tasks involving sensitive materials. The original framework implements an ensemble of Proxy Tasks by combining multiple models and concatenating their feature outputs, resulting in a higher-order vector space for inference. EISP follows the formalization of Proxy Tasks by \citet{Coelho2025MinimizingRisks}, which minimizes risks by using models that are not trained on sensitive data, with only a lightweight ensemble model being trained on vectors extracted from sensitive material and therefore being access-restricted. The system is combined with explainability techniques, such as dimensionality-reduction visualization and feature importance using SHAP values~\cite{lundberg2017unifiedapproachinterpretingmodel}.

In this paper, we introduce improvements to the EISP framework's design and validation protocol and apply it for the first time to real CSAI classification, leveraging a dataset from Brazil's Federal Police, the Region-based annotated Child Pornography Dataset (RCPD)~\cite{Macedo2018BenchmarkMethodologyChild}. To specialize EISP for CSAI, we provide a novel selection of Proxy Tasks comprising seven different models related to the target task, including pose estimation, nudity classification, perceived gender, and age estimation. The extracted features feed an XGBoost model \cite{Chen_2016} as an ensemble, surpassing the state of the art in inference accuracy. The pipeline also includes visualizations and feature importance for each Proxy Task, providing an explainable pipeline for LEA and, in turn, being compliant with new regulations on AI usage, such as the Artificial Intelligence Act (AIA)~\cite{Walke2023AIAct}. 

We additionally evaluate the performance of a baseline representation learning DINO~\cite{caron2021emergingpropertiesselfsupervisedvision} model, using the extracted features with an XGBoost model. Contrasting this method with ours, we show that our Proxy Task ensemble is not only more explainable but also more accurate. Finally, as improvements to the original framework, we added K-Fold Cross-Validation for a robust experimental protocol and improved the dimensionality reduction of extracted features, optimizing the final dimensionality per feature. We made all changes publicly available as a general-purpose Python library for inference on sensitive images\footnote{The source code for our modified implementation of EISP is available at \href{https://github.com/araceli-project/eisp}{github.com/araceli-project/eisp}, and the application of the framework on RCPD is available at \href{https://github.com/araceli-project/EISPCSAI}{github.com/araceli-project/EISPCSAI}.}.

In summary, our contributions are:

\begin{enumerate}
    \item Through applying the EISP framework for the first time on real CSAI, we realize the potential of proxy tasks for this sensitive task and
    \item we demonstrate that such ensemble model is not only performant but is also a path towards more explainability, allowing us to show how each task contributes to a final CSAI classification.
\end{enumerate}

\section{Related Work}

Research on CSAI classification spans multiple directions, including traditional identification of known content, learning-based approaches for novel material, and strategies that rely on sub-tasks under strong data access constraints. In this section, we organize prior work along these axes, covering methods for detecting known and unseen CSAI, approaches based on sub-tasks, efforts toward explainability, and ensemble-based frameworks designed for sensitive domains.

\subsection{Detection of Known and Novel CSAI}

Hash-based methods, such as PhotoDNA~\cite{photodna}, have been widely applied to identify redistribution of known images stored in CSAI databases. However, these are ineffective against new material and can be eluded by little alterations~\citep{Westlake2012ComparingCSAM}; even methods based on perceptual hashes suffer from similar or worse shortcomings, such as the model inversions shown by \citet{Struppek2022LearningBreakDeep}. As an alternative, researchers have invested in computer vision pipelines~\cite{SaeBae2014TowardsAutomaticCSAM} and deep learning-based methods~\cite{Macedo2018BenchmarkMethodologyChild} as promising approaches to identify novel content shared online. Nevertheless, data restrictions in the CSAI domain pose several challenges for training and evaluating machine learning pipelines.  
In this context, learning-based approaches must rely on indirect supervision strategies, motivating the use of alternative tasks and representations.

\subsection{Proxy Task Based Approaches for CSAI}

Early methods for classifying novel CSAI framed the task as nudity classification, measuring the level of skin exposure~\cite{DeCastro2010NudetectiveCSAM, DeCastro2012StatisticalCSAM} or extracting skin-related features~\cite{schulze2014automatic}, with a later work also adding a child classification pipeline~\cite{SaeBae2014TowardsAutomaticCSAM}. Deep learning-based methods later followed a similar protocol, pursuing sub-tasks related to CSAI, such as pornography classification and age estimation~\cite{Macedo2018BenchmarkMethodologyChild, Gangwar2021AttMCNNAttentionMetric}, training on publicly available datasets and testing on LEA apprehensions. The work of \citet{Vitorino2018LeveragingCSAM} is among the few exceptions that have experimented with end-to-end CSAI classification, which required fine-tuning directly on law enforcement datasets, thus producing models that could not be distributed.  

In addition to pornography and age, other substitute targets have been pursued. \citet{laranjeira2022seeinglookinganalysispipeline} proposed an analysis pipeline tailored to CSAI leveraging a wide variety of tasks, including skin tone estimation, object detection, and scene classification. The latter has been later pursued~\cite{valois2025leveraging, Coelho2025MinimizingRisks,coelho2024transformers}, establishing the relevance of contextual cues for CSAI.  
While these approaches leverage related tasks to approximate CSAI classification, they typically operate in isolation and do not explicitly address how to combine multiple signals under practical constraints such as limited access to sensitive data.

\citet{Coelho2025MinimizingRisks} formalized the concept of training models in related sub-tasks as Proxy Tasks. They take into account not only risk minimization through training models in related sub-tasks, as has been common in prior literature, but also additional constraints, such as the lack of specialized infrastructure within LEA, the domain adaptation challenges inherent to training on related domains, and the limited personnel available to curate and annotate highly restricted datasets. This perspective shifts the problem from designing isolated predictors to building systems that can operate under real-world legal and operational constraints.

\subsection{Explainability in CSAI Classification}

An additional concern in the domain of CSAI is explainability~\cite{sanchez2019practitioner}. The work of \citet{ORONOWICZJASKOWIAK2024102619} uses expert knowledge and a deep learning model to achieve explainability by analyzing activation maps of the presence of specific body parts in the image. However, it is limited to body parts analysis and cannot relate to other factors, such as objects or the scene depicted in the image. Furthermore, the work of \citet{barros2025attention} uses Scene Graphs and attention networks to achieve explainability in CSAI classification by producing attention scores for each graph node, but is limited to using only one Proxy Task, namely Scene Graph Generation~\cite{Caetano:SIBGRAPI:2024}. These limitations highlight the need for approaches that combine multiple sources of evidence while preserving interpretability.

\subsection{Ensemble-Based Approaches}

Ensemble methods have been previously studied as a means of enhancing explainability. \citet{10.1145/3292522.3326027} used a collection of scalar Natural Language Processing features for fake news classification. The authors conducted experiments with hundreds of feature combinations, estimating not only inference accuracy but also feature importance and producing visualizations for explainability to identify the best-performing set of features. This solution inspired our previous work~\cite{EISPErnesto2025}, in which we proposed the Ensemble for Inference in Sensitive data using Proxy Tasks (EISP) framework. The design leveraged an ensemble of deep learning-based features to classify images from sensitive domains, providing a distributable and thus reproducible classification pipeline without training on sensitive data. We, however, validate our work on a public dataset rather than on sensitive, private material. Furthermore, the framework's design and validation protocol could be improved.  
In particular, prior work does not fully explore the impact of evaluating such systems in real-world law enforcement datasets nor the effect of systematically combining diverse proxy tasks under a unified evaluation protocol.
\section{Methodology}

\begin{figure*}[t]
    \centering
    \includegraphics[width=0.665\linewidth]{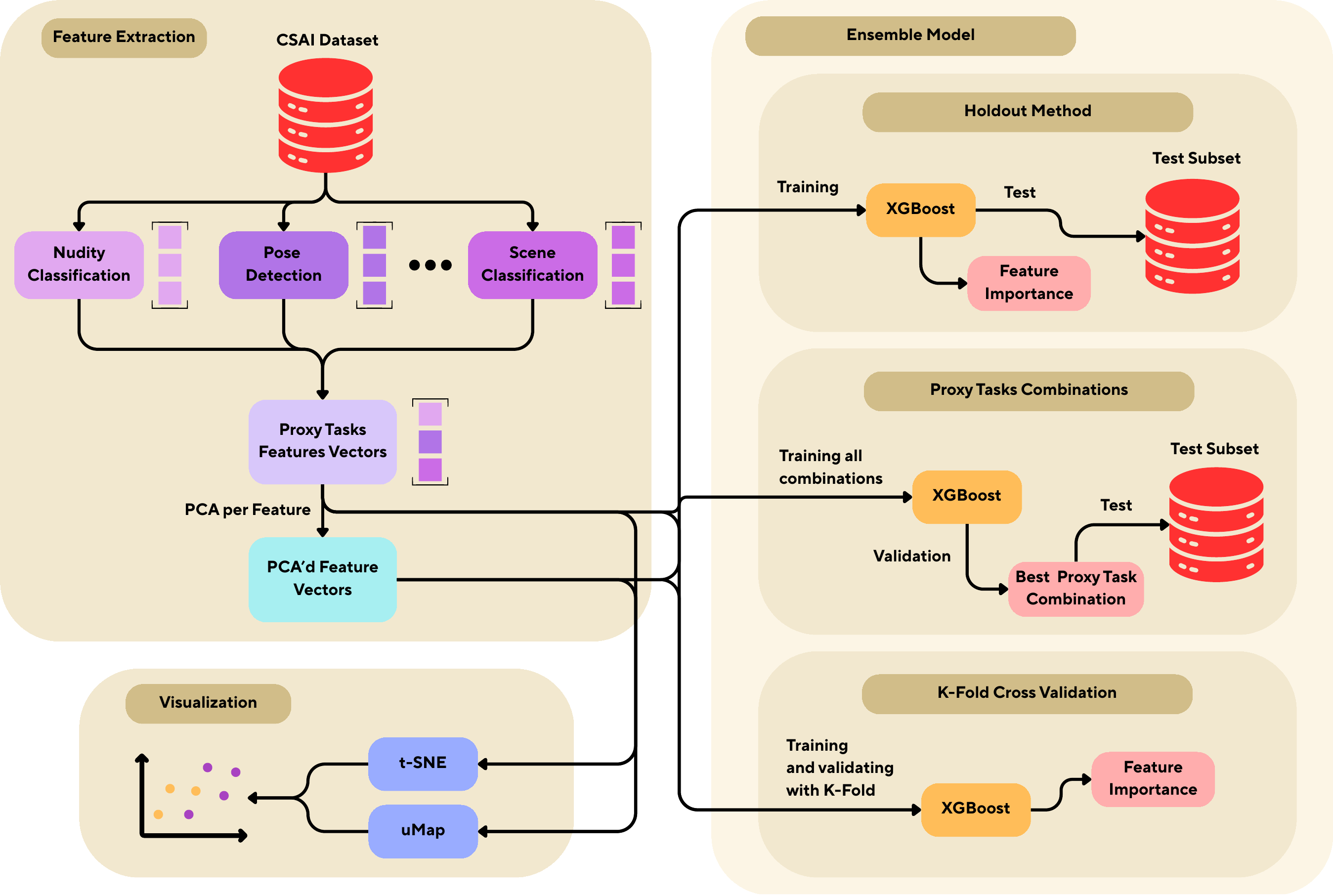}    
    \caption{Diagram of our modified version of the EISP framework, composed of Feature Extraction, Visualization, and Ensemble Model.}
    \label{fig:methodology}
    \Description {
    Diagram of our modified version of the EISP framework, composed of Feature Extraction, Visualization, and Ensemble Model. In Feature Extraction, vectors from the CSAI dataset are processed by several off-the-shelf models, in the image Nudity Classification, Pose Detection and Scene Classification are listed, which produce one feature vector per model. Next, all vectors are concatenated, and an alternative feature space is produced by applying PCA to each feature vector. The concatenated vectors are then used on the Visualization step, to produce both t-SNE and uMap plots. In the Ensemble Model step, the concatenated vectors are used for 3 different experiments. The Holdout Method, Proxy Tasks Combinations and K-Fold Cross Validation. All experiments use a XGBoost model, and the first two apply models to the Test Subset of the CSAI Dataset
    }
\end{figure*}

Our approach builds upon the EISP framework~\cite{EISPErnesto2025} and applies it to CSAI classification; the original framework is structured into three main steps, as shown in Figure~\ref{fig:methodology}: (i) Feature Extraction; (ii)~Visualization of Feature Vectors; and (iii) Ensemble Model. The feature extraction step leverages a curated collection of Proxy Tasks in the CSAI domain, for which off-the-shelf models are available. As detailed later, we have contributed to this step by adding a novel proxy task, modifying the aggregation of demographics-related features, and including an experiment that replaces all proxy tasks with a single state-of-the-art image representation model, allowing us to evaluate the trade-offs between explainable representations provided by Proxy Tasks and high-capacity image features.

The subsequent step, visualization, contributes to explainability by providing insights into how individual features separate the data according to the given labels. Lastly, we propose three settings for training the ensemble model: robust k-fold cross-validation across the entire dataset, exhaustive feature combination experiments (single feature, pairs, groups of three, and so on), and training with the complete set of proxy tasks. The latter two approaches include an essential contribution of our work: splitting RCPD into fixed subsets, such that future work requiring training or fine-tuning on RCPD can be comparable to ours. The data splits were provided to the LEA responsible for RCPD and included in the dataset. Additionally, the released source code includes the procedure to generate these splits.

\subsection{Feature Extraction}

This step includes extracting latent vectors for each selected model and processing PCA for dimensionality reduction. We experiment with both the original and reduced features, since lightweight inferences are preferable for LEA, as long as the quality of results is maintained. Unlike the original EISP framework, which proposes a fixed size for dimensionality reduction, we suggest optimizing PCA processing per feature, establishing a criterion of preserving the explained variance ratio above 80\%.

The curation of proxy tasks in our previous work~\cite{EISPErnesto2025} is inspired by \citet{laranjeira2022seeinglookinganalysispipeline}, including demographic attributes (age, gender, and skin tone), pornography-related concepts (nudity), and contextual cues (objects and scene). Our contribution is three-fold: first, we include pose detection as a novel target, since it has been cited as relevant by LEA~\cite{Yiallourou2017DetectionImagesContaining, kloess2019challenges} but, to the best of our knowledge, has never been explored to classify real CSAI. Secondly, we modify the extraction of demographics-related features. While EISP averages features from all individuals, we use features from the person with the youngest estimated age. This proposition follows indications in \citet{Macedo2018BenchmarkMethodologyChild}, stating that regardless of the source of sexual explicitness within an image, the presence of a child constitutes CSAI. Lastly, we propose an experiment replacing all proxy tasks with a single general-purpose image representation model to compare against our explainable pipeline. The remainder of this subsection describes the off-the-shelf models used in our work.

\subsubsection{DINO Image Representation}
\label{sec:dino_explanation}
We used DINO-vits8~\cite{caron2021emergingpropertiesselfsupervisedvision} as it is among the state of the art for image representation. We chose this version of DINO over newer models due to both LEA's limited hardware constraints and its environmental footprint. We estimated DINOv1's carbon emission with wAIter~\cite{breder_2025_17247278} to be approximately $100$kg of CO2, compared to DINOv2's~\cite{Oquab2023DINOv2LearningRobust} emission of $3.7$ tons and DINOv3's~\cite{simeoni2025dinov3} emission of $18$ tons of CO2. When compared with Proxy Tasks features, accuracy will reveal whether explainable features hinder inference performance.

\subsubsection{Proxy Tasks}
We employed six Proxy Tasks, leveraging state-of-the-art off-the-shelf models for each task. For scene classification, we also experimented with the work of \citet{Coelho2025MinimizingRisks}, which, although not the best-performing method for this task, is tailored to the CSAI domain. Therefore, we used a total of seven models.

\paragraph{Nudity Classification}
We used a pre-trained ViT-based model~\cite{AdamCodd/vit-base-nsfw-detector}, which the authors describe as restrictive, as it classifies certain non-pornographic images, such as cleavage or substantial skin exposure, as NSFW\footnote{Not Safe For Work (NSFW) is usually a synonym to pornography and related media. Throughout this article, NSFW classification will be called nudity classification, due to CSAI being a deeply harmful crime, and not consensual pornography.}. Our feature extraction uses the latent feature from the first classification (CLS) token of the architecture.

\paragraph{Object Detection}
The current state of the art for object detection is YOLOv11x~\cite{Jocher_Ultralytics_YOLO_2023} trained on the COCO dataset. We extract vectors from the last feature layer of the architecture.

\paragraph{Pose Detection}
YOLOv11x~\cite{Jocher_Ultralytics_YOLO_2023} is also trained on MSCOCO for pose detection, with minor architectural adaptations. Similarly to object detection, we extract the last feature vector produced.

\paragraph{Scene Classification}
We used two models, each with a different training method, to compare them. The first one is a pre-trained AlexNet model trained in the Places365 dataset~\cite{zhou2017places}, and the second one is a model made with CSAI constraints in mind, which was trained with few-shot learning and distance-based representation (from \citet{Coelho2025MinimizingRisks}), also being evaluated for indoor scenes, the most common in CSAI according to LEA~\cite{valois2025leveraging}.

\paragraph{Age and Perceived Gender Prediction}
These two targets are treated as a single Proxy Task since we leverage a single model tailored to CSAI~\cite{Macedo2018BenchmarkMethodologyChild}. Their proposed pipeline starts by extracting crops for each visible face using MTCNN~\cite{ivan_de_paz_centeno_2024_13901379}, then performing age estimation with a custom-trained model. We extract the final latent feature from the latter model corresponding to the younger predicted face. For images without detectable faces, we use a zero vector.

\paragraph{ITA Skin Tone Feature}
This is the only feature represented by a scalar value (in the Individual Typology Angle scale). Its calculation requires extracting face crops and localizing facial landmarks~\cite{merler2019diversityfaces}. With the corresponding pixels, the metric~\cite{10.1007/978-3-030-59725-2_31} is calculated as:
$$ITA = \frac{\arctan(L-50)}{b} \times \frac{180}{\pi},$$
$L$ and $b$ are image channels representing luminance and yellow amount, respectively.
If there are no recognizable faces, the value is $0$ instead.

\subsection{Visualization of Feature Vectors}
We used t-SNE~\cite{Maaten08tsne} and UMAP~\cite{mcinnes2020umapuniformmanifoldapproximation} as dimensionality reduction techniques to produce scatter plots, to visualize clusters and patterns in the feature space. Specifically, we plot the reduced representation of features from each Proxy Task, as well as the concatenation of all features, to visually assess how each model and their combination separate the data by the target label. Additionally, we produced a visualization that associates each point with its corresponding image, enabling our LEA partners to provide qualitative insights into the model's behavior. This plot cannot be distributed as it contains CSAI, but the LEA's qualitative assessment is detailed later.

\subsection{Ensemble Model}

In this stage, we use the extracted features to train a lightweight ensemble model targeting binary CSAI classification. Following \citet{10.1145/3292522.3326027}, we train a high-performance gradient boosting algorithm, called XGBoost~\cite{Chen_2016}, capable of combining the proposed set of features into a strong, iteratively optimized model. \citet{10.1145/3292522.3326027} also proposed the extraction of Shapley Additive Explanations~\cite{lundberg2017unifiedapproachinterpretingmodel}, known as SHAP values, which estimate how each feature contributes to a model's prediction. In this paper, we use SHAP values to understand the overall behavior of our models, but they may also be included in a final application to provide explanations for individual predictions.

Since RCPD was designed for testing purposes, it did not provide data splits. Leveraging prior knowledge from Proxy Tasks, our ensemble model trained for CSAI classification can learn from a small number of samples, enabling robust experimental protocols with training, validation, and testing. By contacting the LEA responsible for RCPD, we proposed fixed data splits so that future methods requiring training or fine-tuning on RCPD can benchmark their results against ours. We balanced our proposed splits with respect to RCPD's definition of CSAI within an image: the presence of a child up to 13 years old and depictions of nudity or sexual activity. The remainder of this subsection details three experiment protocols we conducted using both original and PCA-processed features. 

\subsubsection{Holdout Method} 
Using the full set of features across all Proxy Tasks in our work, we tune XGBoost hyperparameters (learning rate, max. depth, subsample ratio and column subsample ratio) on the training and validation splits. The resulting model is then applied to the test set. Afterwards, we extract SHAP values from the training split, aggregating them into the average feature importance of each Proxy Task.

\subsubsection{Combination of Proxy Tasks}
With the hyperparameters optimized through the Holdout Method, we exhaust all possible combinations of Proxy Tasks using the training and validation sets. The goal is to identify which models can be dropped to optimize the system's storage, training, and inference compute resources, while maintaining accuracy.
We applied the ensemble model corresponding to the best combination of Proxy Tasks on the test subset of RCPD to compare with the model produced through the Holdout~experiment.

\subsubsection{K-Fold Cross-Validation}
We used the average balanced accuracy across five folds of the training subset to tune the hyperparameters of the XGBoost model for the validation of using all Proxy Tasks models.
For each iteration, we also extracted SHAP values and aggregated them to calculate feature importance values for each Proxy Task, finally producing a bar plot of feature importance.

\section{Experiments and Results}

\subsection{RCPD Dataset}
We evaluated our method on the Region-based annotated Child Pornography Dataset (RCPD)~\cite{Macedo2018BenchmarkMethodologyChild}, as it is one of the very few CSAI benchmarks available for testing\footnote{Evaluation instructions may be retrieved from \url{https://patreo.dcc.ufmg.br/2019/03/09/region-based-annotated-child-pornography-dataset}}. The dataset comprises $2,138$ samples, including CSAI, adult pornography, and safe images. It is annotated with hierarchically organized bounding boxes for person and body parts, along with class labels for perceived demographic attributes (age and gender), and nudity level (none, seminude, nude, and sexual interaction). From the provided labels, the authors derive a binary CSAI classification target as images simultaneously depicting children up to 13 years old and some instance of nudity or sexual interaction. The dataset contains $836$ CSAI instances, $286$ adult pornography, $508$ safe images of people, 
and the remainder being safe images without people, such as landscapes.

Our work uses only the image itself and the CSAI binary label; all other annotations were disregarded, as the objective was to experiment with feature vectors and improve reproducibility on other datasets. We also reserved a fixed test subset comprising 20\% of the images, as there was no predefined test set in~RCPD. 

RCPD was originally produced in partnership with the Federal Police of Brazil and remains under their custody. Naturally, all data processing and access to the dataset were performed exclusively through the responsible LEA.

\subsubsection{Dataset Limitations} \label{sec:rcpd-limitations}
Despite its relevance as a real-world dataset, RCPD presents important limitations that must be considered when interpreting the results. Access to the data is highly restricted, preventing direct experimentation by researchers and requiring all evaluations to be conducted within law enforcement environments. Moreover, the dataset was not designed as a standard machine learning benchmark, lacking predefined splits and reflecting operational conditions rather than curated experimental settings, which introduces challenges for reproducibility and controlled comparisons across methods. In addition, prior analyses have identified biases in its composition~\cite{laranjeira2022seeinglookinganalysispipeline}, such as the predominance of lighter-skinned individuals, which does not reflect real Child Sexual Abuse statistics in Brazil. The dataset also presents imbalances across relevant categories, with $836$ CSAI samples compared to approximately $400$ safe images containing children and only $286$ instances of adult pornography~\cite{macedo2025child}. These characteristics may affect generalization, for instance, limiting performance on more diverse populations or favoring features such as nudity detection, which alone can yield strong results.

\subsection{Visualization of Feature Vectors}

\begin{figure*}[tp]
     \centering
     \begin{subfigure}[b]{0.435\textwidth}
         \centering
         \includegraphics[width=\linewidth,trim={2.6cm 2cm 2.1cm 2cm},clip]{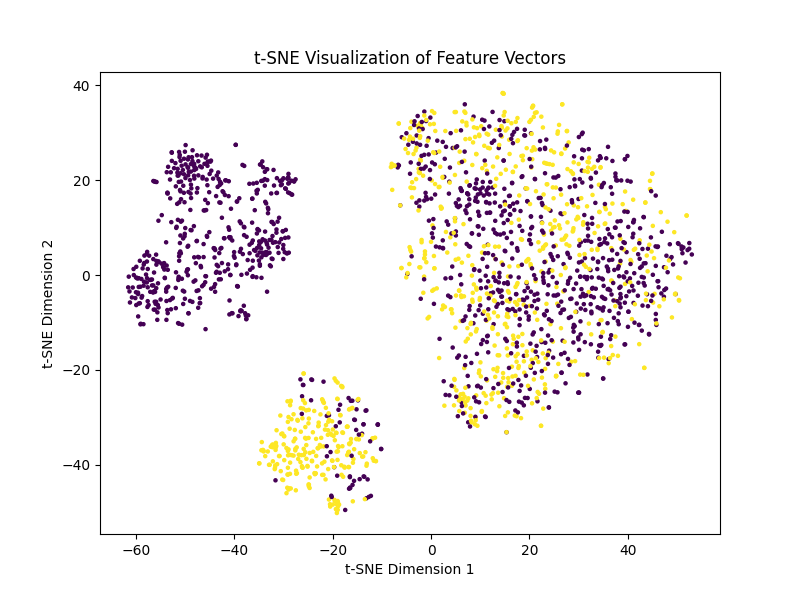}
         \caption{t-SNE}
         \label{fig:tsne_all}
     \end{subfigure}
     \hfill
     \begin{subfigure}[b]{0.435\textwidth}
         \centering
         \includegraphics[width=\linewidth,trim={2.7cm 2cm 2.1cm 2cm},clip]{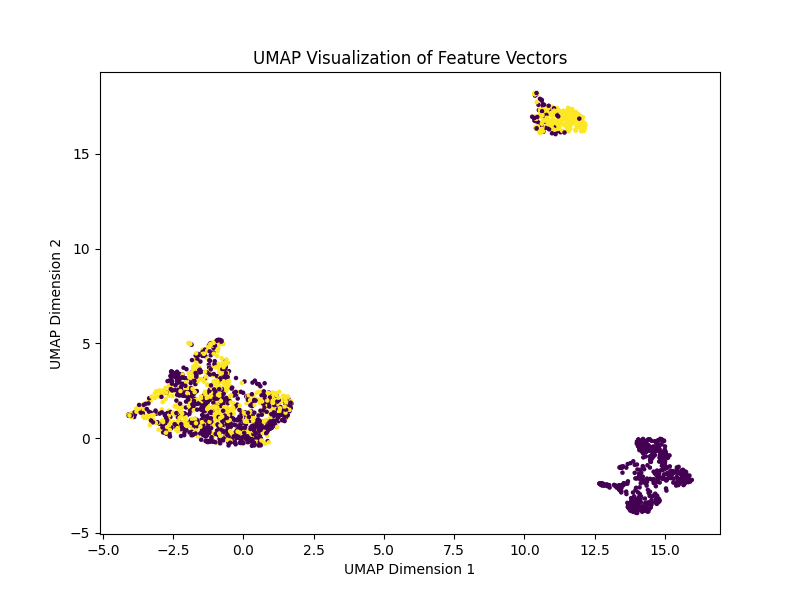} 
         \caption{UMAP}
         \label{fig:umap_all}
     \end{subfigure}
     \caption{2D projections of all concatenated feature vectors after PCA-processing. Yellow dots represent CSAI images and purple dots represent Non-CSAI images. Three clusters are visible: one showing mixed images, one mostly composed of CSAI, and another composed entirely of non-CSAI.}
     \Description{2D projections of all concatenated feature vectors after PCA-processing. Yellow dots represent CSAI images and purple dots represent Non-CSAI images. Three clusters are visible: one showing mixed images, one mostly composed of CSAI, and another composed entirely of non-CSAI. There is a cluster made of purple dots, a cluster made mostly of yellow dots, and a mixed cluster.}
    \label{fig:tsne_umap_all}
\end{figure*}

\begin{figure*}[tp]
     \centering
     \begin{subfigure}[b]{0.47\textwidth}
         \centering
         \includegraphics[width=\linewidth,trim={2.6cm 2cm 2.1cm 2cm},clip]{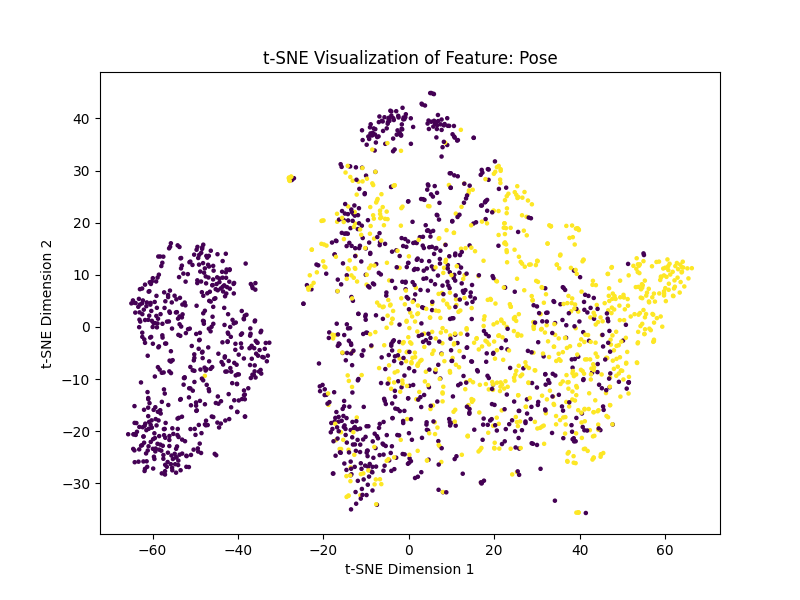}
         \caption{Pose detection}
         \label{fig:tsne_pose}
     \end{subfigure}
     \hfill
     \begin{subfigure}[b]{0.47\textwidth}
         \centering
          \includegraphics[width=\linewidth,trim={2.6cm 2cm 2.1cm 2cm},clip]{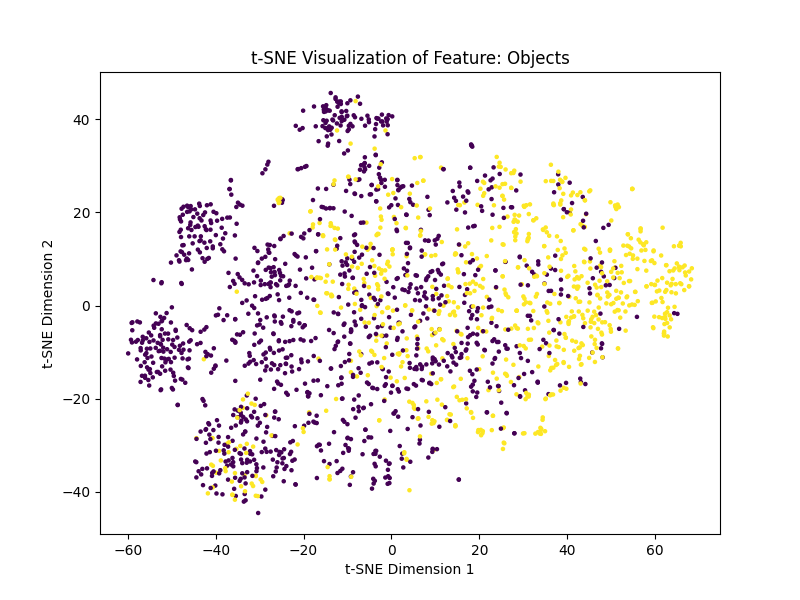}
         \caption{Object detection}
         \label{fig:tsne_object}
     \end{subfigure}
     \hfill
     \begin{subfigure}[b]{0.47\textwidth}
         \centering
         \includegraphics[width=\linewidth,trim={2.6cm 2cm 2.1cm 2cm},clip]{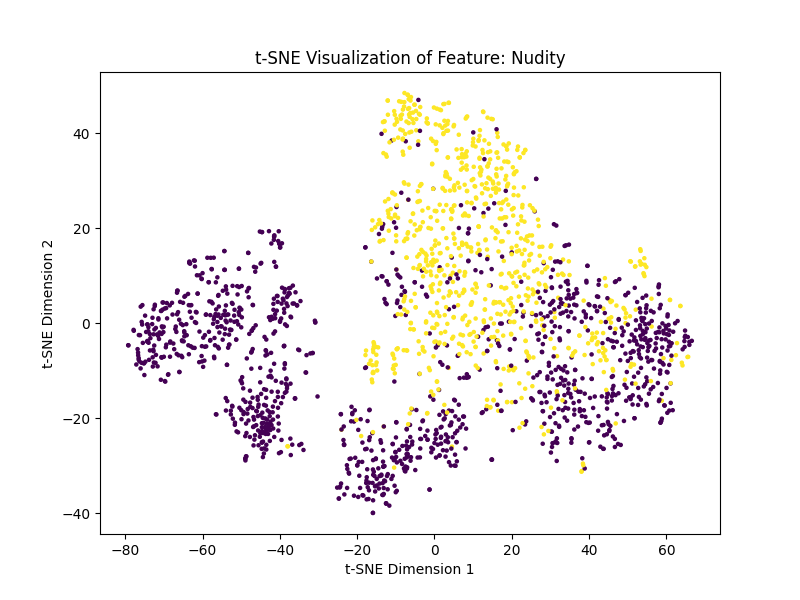} 
         \caption{Nudity classification}
         \label{fig:tsne_nudity}
     \end{subfigure}
     \hfill
     \begin{subfigure}[b]{0.47\textwidth}
         \centering
         \includegraphics[width=\linewidth,trim={2.6cm 2cm 2.1cm 2cm},clip]{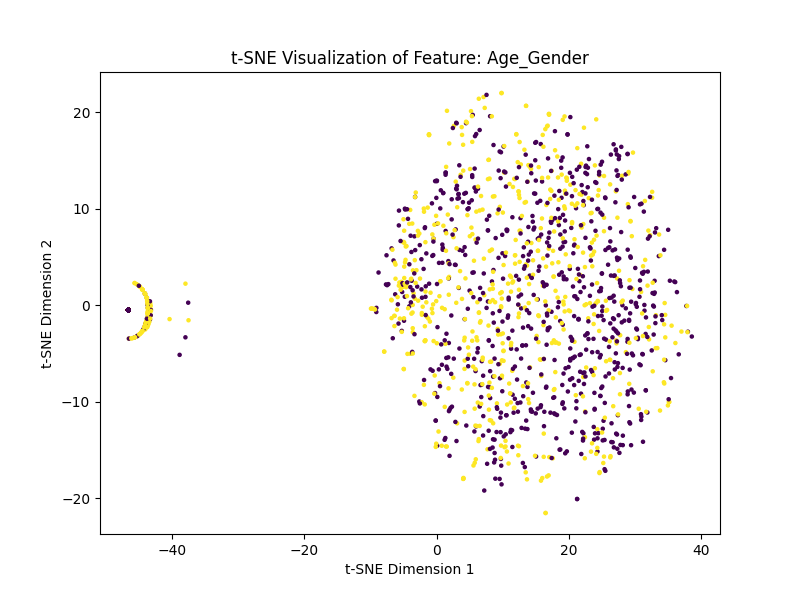}
         \caption{Age and perceived gender recognition}
         \label{fig:tsne_agegender}
     \end{subfigure}
     \hfill
     \begin{subfigure}[b]{0.47\textwidth}
         \centering
         \includegraphics[width=\linewidth,trim={2.6cm 2cm 2.1cm 2cm},clip]{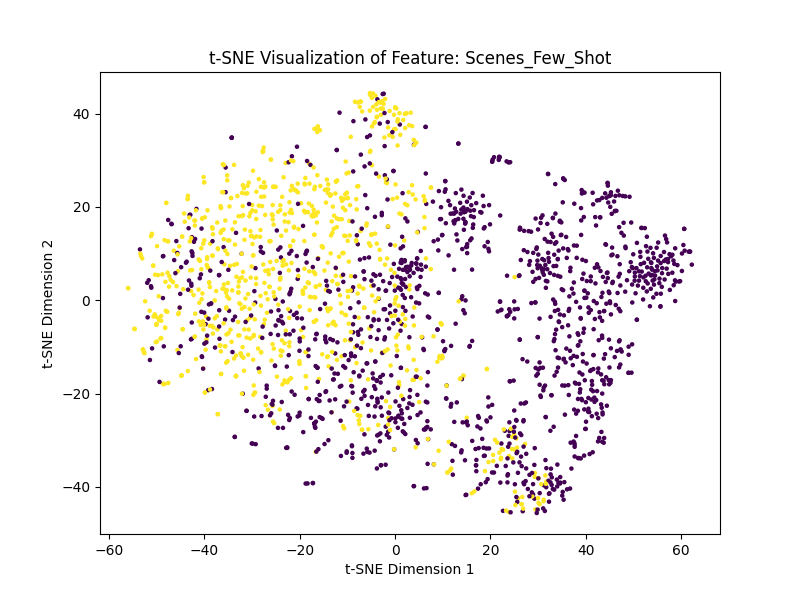}
         \caption{Few-shot scene classification}
         \label{fig:tsne_fewshot}
     \end{subfigure}
     \hfill
     \begin{subfigure}[b]{0.47\textwidth}
         \centering
         \includegraphics[width=\linewidth,trim={2.6cm 2cm 2.1cm 2cm},clip]{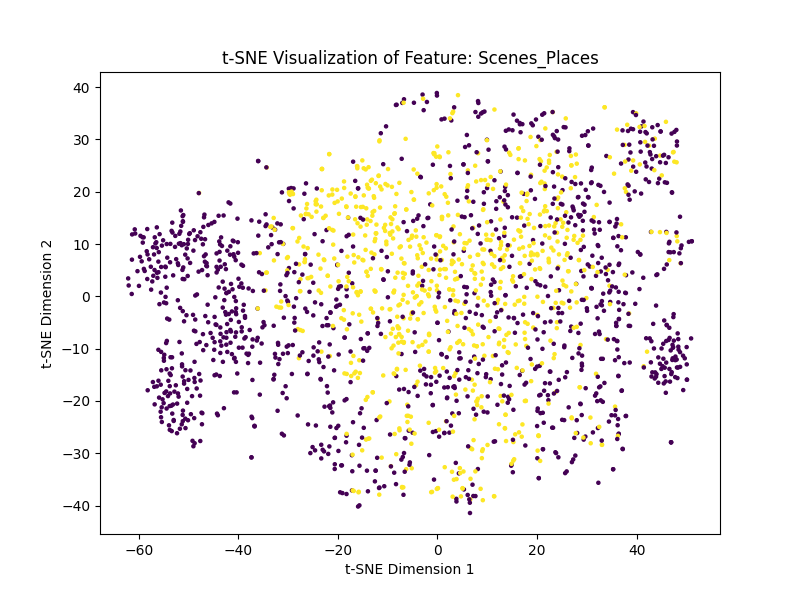}
         \caption{Places scene classification}
         \label{fig:tsne_places}
     \end{subfigure}
     \caption{t-SNE projections of the feature space for each Proxy Task model after PCA preprocessing. Yellow dots represent CSAI images and purple dots represent Non-CSAI images.}
     \Description{t-SNE projections of the feature space for each Proxy Task model after PCA preprocessing. Yellow dots represent CSAI images and purple dots represent Non-CSAI images. Most plots show a clear division between yellow and purple dots, except the Age and perceived gender model.}
    \label{fig:tsne_features}
\end{figure*}

When observing both the t-SNE (Figure~\ref{fig:tsne_all}) and UMAP (Figure~\ref{fig:umap_all}) plots of all concatenated Proxy Tasks features, there are three major clusters and a clear perceived separation by target label. The law enforcement agent responsible for RCPD analyzed these clusters. The cluster that has mixed labels contains images with easy to perceive faces, the cluster that contains mostly CSAI (yellow dots) is comprised of images without or with hard to detect faces --- a cluster that is also present in the plot extracted from the Age/Gender model (Figure~\ref{fig:tsne_agegender}) --- and the last one is comprised entirely of non-CSAI without people, which can also be clearly seen on the plots extracted from Pose (Figure~\ref{fig:tsne_pose}) and Nudity (Figure~\ref{fig:tsne_nudity}) features.

We observed a clear separation of data by the target label in most Proxy Tasks and DINO features, both on t-SNE (Figure \ref{fig:tsne_features}) and UMAP (Figure \ref{fig:umap_features}) plots. The UMAP plots for each model show similar behavior to that of the t-SNE plots and are shown in Appendix~\ref{sec:umap_per_model}. Comparing these Proxy Tasks with the DINO visualization, both achieved similar perceived separation and can be used by LEA together with a scatter plot with image markers to explore clusters and identify which images are harder for the model to discern. The advantage of using the Proxy Tasks model plots is that they provide initial explainability of how the data is being separated, as the models were trained for a specific task. Naturally, the ITA Skin Tone Proxy Task could not generate a scatter plot, as it is represented by a single value rather than a high-dimensional feature.

Lastly, the qualitative analysis of t-SNE plots from Figure~\ref{fig:tsne_features} associated with the corresponding images found the Nudity classification scatter plot as the most interesting, highlighting the presence of clusters separated by gender, nudity levels, pornography, and even a dense cluster dedicated to artificial images such as cartoons and advertisements. Concerningly, the model placed young girls in swimwear near adult pornography, while boys in such attire were more distant. A final remark regarding nudity features were aggregations of different types of child sexual abuse, although they did not elaborate any further. 

The LEA expert also highlighted helpful clusters derived from scene-related features, including close-ups with few visible contextual cues, images without people, and specific scene clusters such as beaches, bathtubs, bedrooms, and ``green areas'' (outdoor vegetation). 
Furthermore, the Age/Gender model did not cluster consistently by age nor gender, but images of the same individual often appeared together. For the remaining clusters, namely object and pose, the clusters were semantically consistent with the corresponding target, with no further remarks.

\begin{figure}[t]
    \centering
    \includegraphics[width=0.6\linewidth,trim={2.7cm 2cm 2.1cm 2cm},clip]{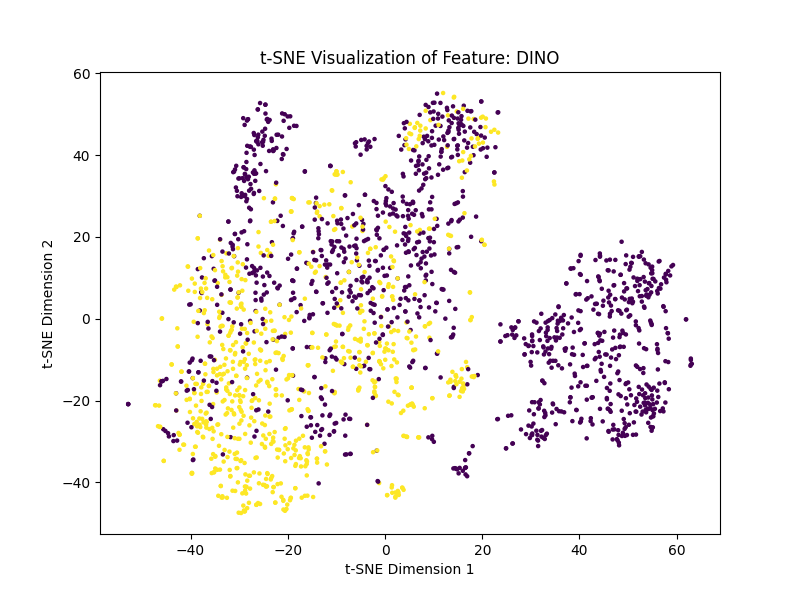}
    \caption{t-SNE projection of DINO's feature space after \text{PCA-processing}. Yellow dots represent CSAI images and purple dots represent Non-CSAI images.}
    \Description{t-SNE projection of DINO's feature space after \text{PCA-processing}. Yellow dots represent CSAI images and purple dots represent Non-CSAI images. The plot show a clear division between yellow and purple dots.}
    \label{fig:tsne_dino}
\end{figure}

\subsection{Ensemble Experiments}

We compared all experiments with their PCA-processed versions for each Proxy Task model and DINO feature vectors. The original dimension size and the reduced dimension size of each feature are presented in Table~\ref{tab:dimensionsize}. Maintaining at least 80\% of explained variance ratio, PCA achieved a reduction to $10.7\%$ of the original dimension size of the concatenated features, and a reduction to $33.7\%$ of the DINO features dimension size.

\begin{table}[t]
\small
\caption{Dimension size for each model feature space before and after PCA application.\vspace{-0.2cm}}
\begin{tabular}{lrr}
\midrule
\multicolumn{1}{l}{}           & \multicolumn{1}{l}{\textbf{Original}} & \multicolumn{1}{l}{\textbf{PCA}} \\ \midrule
Pose                             & 768                           & 16                       \\ 
Object                           & 768                           & 32                       \\ 
Nudity                           & 768                           & 256                      \\ 
Age/Gender                       & 4096                          & 256                      \\ 
ITA                              & 1                             & 1                        \\ 
Scene Places                     & 256                           & 64                       \\ 
Scene Few Shot                   & 384                           & 128                      \\ 
Concatenated                     & 7041                          & 753                      \\ 
DINO                             & 384                           & 128                      \\ \midrule
\end{tabular}
\label{tab:dimensionsize}
\end{table}

\subsubsection{K-Fold Cross-Validation}
This experiment used Optuna~\cite{optuna_2019} with $5$~optimization steps to tune hyperparameters. With the best hyperparameter, our model achieved an average balanced accuracy across folds of $89.9\%$ and recall of $87.1\%$ with the original features and average balanced accuracy of $90.0\%$ and recall of $87.5\%$ with PCA-processed features using all Proxy Tasks. When using only DINO features, the best hyperparameter achieved $86.3\%$ and $82.9\%$ on the original features and $87.0\%$ and $85.5\%$ on the reduced features, using the same metrics. As such, the Proxy Tasks achieved better results than the non-explainable DINO features on both the original and reduced features. Table~\ref{tab:kfold_bal_acc} shows the accuracy per fold.

\begin{table*}[h]
\small
\caption{Balanced accuracy per fold, average balanced accuracy, average recall, average precision and standard deviation (all values in \%), using original features and PCA'd features with best hyperparameters, on both Proxy Tasks and DINO features.\vspace{-0.2cm}}
\begin{tabular}{lcccccccc}
\midrule
         & \textbf{Fold 1} & \textbf{Fold 2} & \textbf{Fold 3} & \textbf{Fold 4} & \textbf{Fold 5} & \textbf{Avg. Bal. Acc.} & \textbf{Avg. Recall} & \textbf{Avg. Precision} \\ \midrule
Orig. P.T. & 90.8 & 86.4 & 89.3 & 92.8 & 90.1 & \textbf{89.9} \tiny{$\pm~2.3$} & \textbf{87.1} \tiny{$\pm~2.1$} & \textbf{88.6} \tiny{$\pm~3.1$} \\ 
Orig. DINO & 89.7 & 83.7 & 87.6 & 87.0 & 83.6 & 86.3  \tiny{$\pm~2.6$} & 82.9 \tiny{$\pm~4.0$} & 84.3 \tiny{$\pm~1.8$} \\ 
PCA P.T.      & 90.4 & 88.5 & 91.2 & 91.4 & 88.3 & \textbf{90.0}  \tiny{$\pm~1.4$} & \textbf{87.5} \tiny{$\pm~1.9$} & \textbf{88.6} \tiny{$\pm~2.0$} \\ 
PCA DINO      & 88.4 & 85.8 & 89.3 & 86.5 & 85.2 & 87.0  \tiny{$\pm~1.8$} & 85.5 \tiny{$\pm~3.1$} & 83.3 \tiny{$\pm~1.4$} \\ \midrule
\end{tabular}
\label{tab:kfold_bal_acc}
\end{table*}

\begin{figure}[ht]
    \centering
    \includegraphics[width=0.925\linewidth]{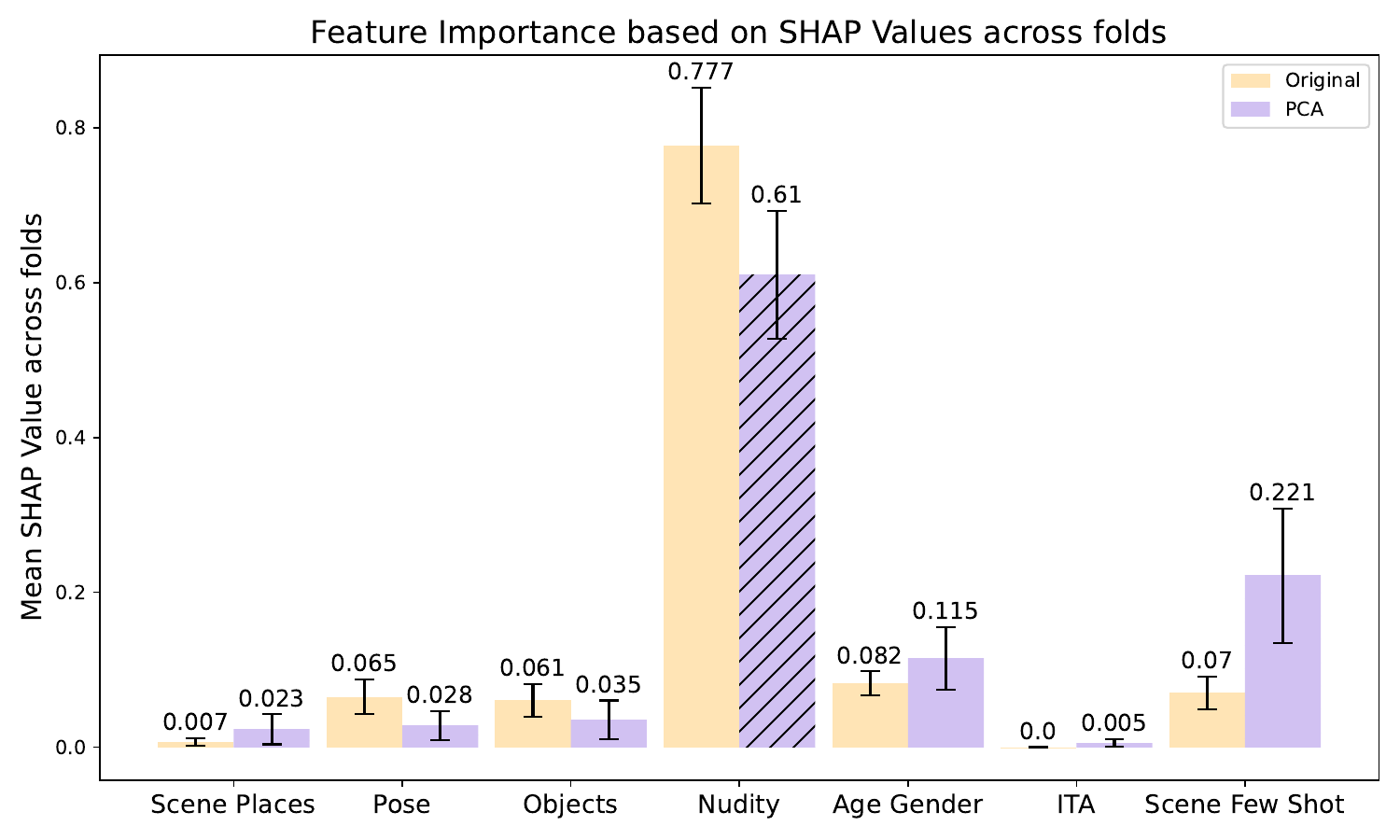}
    \caption{Aggregated SHAP values plot of average feature importance across folds, using both Original and PCA-processed features. Error bars represent standard deviation.}
    \Description{Aggregated SHAP values plot of average feature importance across folds, using both Original and PCA-processed features. Error bars represent standard deviation. The models with higher feature importance are the Nudity, Scene Few Shot and Age Gender models.}
    \label{fig:kfold_featimp}
\end{figure}

As seen in Figure~\ref{fig:kfold_featimp}, the average feature importance values across folds showed greater importance attributed to the Nudity classification, Age estimation and Perceived Gender and Few Shot Scene classification models; both on original and PCA-processed features. The Few-Shot Scene classification model had a significantly greater influence than the model trained on Places. When comparing the original features to PCA-processed features, the Pose estimation, Object detection and Nudity classification models were attributed reduced feature importance values, while other models showed increased values after dimensionality reduction. Finally, the ITA Skin Tone value has a feature importance value near $0$, which could have a higher impact if presented as a feature vector from a deep learning model rather than a scalar value.

\subsubsection{Holdout Method}
This experiment also used Optuna with $5$~optimization steps to tune hyperparameters. With the optimized hyperparameters, the ensemble is trained and tested on the splits fixed by our work. In Table~\ref{tab:results_rcpd_test} we consolidate all results on the test split, including the holdout method leveraging all proxy tasks, the version using only DINO features, and the best model from the Proxy Tasks Combinations experiment (see Section~\ref{sec:exp-proxy-combinations}). In terms of accuracy, the results using all proxy tasks are competitive with the best DINO-based model. Conversely, our approach yields superior precision scores, while PCA-processed DINO features resulted in the best recall on the test split.  
Regarding feature importance for the holdout method, the plots depicted in Figure~\ref{fig:simple_featimp} in Appendix~\ref{sec:crossval_featimp} show similar behavior to the K-Fold plots.

\begin{table}[t]
\caption{Balanced accuracy, recall and precision of each model from this study applied to the test subset of RCPD.\vspace{-0.2cm}}
\small
\begin{tabular}{lccc}
\midrule

 & \textbf{Bal. Acc.} & \textbf{Recall} & \textbf{Precision} \\ 
\textbf{Model} & \textbf{(\%)} & \textbf{(\%)} & \textbf{(\%)} \\ \midrule
DINO + XGBoost Original       & 81.6    &   77.6   & 75.6      \\ 
DINO + XGBoost PCA     & 86.1    &    \textbf{86.5}  & 77.5    \\ 
EISP All Proxy Tasks Original  & 85.6    &    80.8 &   \textbf{82.9}      \\ 
EISP All Proxy Tasks PCA      & 84.5      &       78.2  & \textbf{83.0}      \\ 
EISP Best Combination Original& 85.9       &  81.4   & \textbf{83.0}      \\ 
EISP Best Combination PCA     & \textbf{87.0}  &  84.0  & \textbf{82.9}      \\ 
\midrule
\end{tabular}
\label{tab:results_rcpd_test}
\end{table}

\subsubsection{Proxy Tasks Combinations} \label{sec:exp-proxy-combinations}
We have tested all Proxy Tasks Combinations using the same hyperparameters as the Holdout Method and present the five best combinations (based on balanced accuracy on validation) on Tables~\ref{tab:bestcomb_orig} (for original features) and \ref{tab:bestcomb_pca} (for PCA-processed features). The recall and precision on validation of the best combination using original features are $90.4\%$ and $86.6\%$, and using PCA-processed features are $91.2\%$ and $89.2\%$. The Nudity classification model had the highest frequency across all five best combinations of both features, followed by the Pose estimation, Few-Shot Scene classification and Object detection models. Finally, the combinations using PCA'd features showed slightly higher validation metrics than those using original features.

\begin{table}[h]
\small
\setlength{\tabcolsep}{2.75pt}
\caption{Five best combinations and their respective balanced accuracy (values in \%) on validation using original features.\vspace{-0.2cm}} 
\begin{tabular}{ccccccccc}
\midrule
 & \textbf{Nudity} & \textbf{Pose} & \textbf{Obj.} & \textbf{\makecell{Age/\\Gender}} & \textbf{\makecell{Few\\Shot\\Scene}} & \textbf{\makecell{Places\\Scene}} & \textbf{\makecell{ITA\\Skin\\Tone}} & \textbf{\makecell{Val\\Metric}} \\ \midrule
1° & $\bullet$      &  $\bullet$   &       &  $\bullet$      &                &            &               & \textbf{90.6}     \\ 
2° & $\bullet$      &    & $\bullet$     &          &               &             &            & 90.4     \\ 
3° & $\bullet$      &     &        &   $\bullet$         &             &  $\bullet$            &             & 90.2    \\ 
3° & $\bullet$      &  $\bullet$  &   $\bullet$    &  $\bullet$         &                &              &             & 90.2    \\ 
3° & $\bullet$      & $\bullet$     &     $\bullet$     &     $\bullet$         &      $\bullet$            &              &       $\bullet$        & 90.2     \\ \midrule
\end{tabular}
\label{tab:bestcomb_orig}
\end{table}

\begin{table}[h]
\small
\setlength{\tabcolsep}{2.75pt}
\caption{Five best combinations and their respective balanced accuracy (values in \%) on validation using PCA'd features.\vspace{-0.2cm}}
\begin{tabular}{ccccccccc}
\midrule
 & \textbf{Nudity} & \textbf{Pose} & \textbf{Obj.} & \textbf{\makecell{Age/\\Gender}} & \textbf{\makecell{Few\\Shot\\Scene}} & \textbf{\makecell{Places\\Scene}} & \textbf{\makecell{ITA\\Skin\\Tone}} & \textbf{\makecell{Val\\Metric}} \\ \midrule
1° & $\bullet$      &  $\bullet$   &  $\bullet$     &            &      $\bullet$        &      $\bullet$        &   $\bullet$             & \textbf{91.9}     \\ 
2° & $\bullet$      &  $\bullet$   &    $\bullet$    &    &      $\bullet$          &             &    $\bullet$          & 91.7     \\
2° & $\bullet$       &  $\bullet$   &     &           &  $\bullet$           &              &              & 91.7     \\
3° & $\bullet$      &  $\bullet$    &      &       & $\bullet$            &              &      $\bullet$        & 91.3    \\ 
4° & $\bullet$      & $\bullet$     &      &       &            & $\bullet$            &            & 91.2     \\ \midrule
\end{tabular}
\label{tab:bestcomb_pca}
\end{table}

Applying the best combination of Proxy Tasks on the test subset of RCPD yields $87\%$ balanced accuracy -- our best result on this data split -- and overall shows superior results relative to using all features, as shown in Table~\ref{tab:results_rcpd_test}. Relative to DINO-based results, it shows superior accuracy and precision, although it is slightly behind on recall.

We also report the performance of applying each proxy task individually (Table~\ref{tab:singular_models}). Models leveraging pose, objects, few-shot scene, and nudity achieved over $80\%$ balanced accuracy, with nudity standing out with $90.5\%$ using PCA-processed features. We have previously discussed biases in RCPD that might incur in such a high performance achieved solely with nudity features (see Section~\ref{sec:rcpd-limitations}). 

\begin{table}[t]
\caption{Balanced accuracy of validation using only one model, both using original and PCA'd features.\vspace{-0.2cm}}
\label{tab:singular_models}
\small
\begin{tabular}{lcc}
\midrule
               & \textbf{Original Features} & \textbf{PCA Features}  \\ \midrule
Scene Places   & 77.8              & 75.3          \\ 
Pose           & 82.7              & 82.3          \\ 
Object         & 83.6              & 82.9          \\ 
Nudity         & \textbf{87.1}     & \textbf{90.5} \\ 
Age/Gender     & 59.4              & 52.1          \\ 
ITA Skin Tone  & 52.2              & 50.2          \\ 
Scene Few Shot & 81.9              & 85.7          \\ \midrule
\end{tabular}
\end{table}

\subsubsection{Comparing with the state of the art}
Since RCPD did not have fixed data splits prior to this work, we compare our results on the K-Fold Cross-Validation experiments to metrics provided by other works tackling CSAI classification on RCPD (Table \ref{tab:results_rcpd_cross_val}). We have found the Cross-Validation experiments to be well suited for comparison, since other methods either test on the entire dataset~\cite{DeCastro2010NudetectiveCSAM, Macedo2018BenchmarkMethodologyChild}, conduct similar protocols splitting the data multiple times for fine tuning~\cite{barros2025attention} or make use of few-shot learning protocols~\cite{Coelho2025MinimizingRisks}. In this setting, our ensemble system achieved superior measures of accuracy and precision, with higher recall than almost all methods except \citet{Coelho2025MinimizingRisks}.

\begin{table*}[h]
\caption{Metric and value for other studies and this work, K-Fold experiments applied to the RCPD dataset. $\dagger$ indicates results as reported in \citet{Macedo2018BenchmarkMethodologyChild}.\vspace{-0.2cm}}
\small
\setlength{\tabcolsep}{4pt}
\begin{tabular}{llccc}
\midrule
      \textbf{Work}                         & \textbf{Acc. Metric}    & \textbf{Acc. Value (\%)} & \textbf{Recall (\%)} & \textbf{Precision (\%)} \\ \midrule

NuDetective~\cite{DeCastro2010NudetectiveCSAM}$^\dagger$        & Accuracy & 57.4 & 41.2 &    78.7            \\ 
LED$^\dagger$        
&  Accuracy & 76.5 & 57.2 &  75.3              \\ 
\citet{Macedo2018BenchmarkMethodologyChild}         & Accuracy      & 79.8 & 64.6 &     68.6            \\ 
\citet{barros2025attention}                   & Balanced Accuracy & 74.3 & 76.6 & N/A               \\ 
\citet{Coelho2025MinimizingRisks}                   & Accuracy      & 65.4  & \textbf{91.0} & 60.6                \\ 
EISP All Proxy Tasks Original K-Fold & Balanced Accuracy      & 89.9    & 87.1  & \textbf{88.6}              \\ 
EISP All Proxy Tasks PCA K-Fold & Balanced Accuracy      & \textbf{90.0}  & 87.5   & \textbf{88.6}              \\ 
DINO + XGBoost Original K-Fold & Balanced Accuracy      & 86.3    & 82.9 & 84.3               \\ 
DINO + XGBoost PCA K-Fold & Balanced Accuracy      & 87.0     & 85.5   & 83.3            \\ \midrule
\end{tabular}
\label{tab:results_rcpd_cross_val}
\end{table*}

For the sake of thoroughness, we should mention that the studies from ~\citet{ORONOWICZJASKOWIAK2024102619}, \citet{CastrillnSantana2017EvaluationOL} and \citet{Gangwar2021AttMCNNAttentionMetric} have each achieved accuracy metrics of $93.1\%$, $88\%$ and $92.7\%$ on their own datasets and specific CSAI benchmarks. We are not able to compare these results between each other and against ours due to dataset access restrictions but mention them in the interest of showing our own $91.9\%$ of balanced accuracy is well aligned with what is currently achievable on CSAI or CSAI-related classification tasks.

\subsection{Environmental Footprint}
\label{sec:env_footprint}
We used the CodeCarbon library~\cite{lottick2019energyusagereportsenvironmental} to run all experiments and evaluate energy consumption and carbon emission. It resulted in the energy consumption of $0.274$ kWh and carbon emission of $0.027$~kg.
\section{Discussion}

The use of an ensemble of Proxy Tasks surpassed the state of the art for CSAI Classification using the RCPD dataset. Compared to DINO, it shows that explainable features usage does not hinder inference performance, offers different forms of explainability, and is almost entirely distributable --- except for the lightweight ensemble model trained on the target domain --- making it a viable solution for ethical systems in this~task.

It was also shown to be able to optimize for running on limited hardware, which is needed for several LEAs, both by using PCA'd features and dropping Proxy Tasks through evaluation of their combinations, achieving a light and efficient system. The combinatorial technique has drawbacks: evaluating all Proxy Tasks combinations is time- and compute-intensive, and, as a consequence, its usage needs to be weighed in each scenario. 

Additionally, the Nudity classification task appeared on all experiments as a valuable Proxy Task for CSAI classification, being followed by models trained with CSAI constraints in mind, such as the Age/Gender and Few-Shot Scene models. 

Another observable trend on the results are that models based on PCA processed features overall achieved higher balanced accuracies and Recall scores then models based on original features. We hypothesize that PCA dimensionality reduction can produce a vector space which is easier for the XGBoost model to explore and classify, even as it loses information, as it is a light model.

Finally, the most crucial outcome is that explainable and distributable systems that achieve high accuracy in CSAI classification are viable. In the current context of this task, where regulations, such as European Union's AIA, start to enforce the needed explainability in artificial intelligence systems, and most production-ready systems --- both from LEA and private companies --- are neither explainable nor distributable, the argument that systems with these qualities can be created should guide architects and researchers from new ethical systems to seek both as much reproducibility and different manners of explainability as possible.
\section{Limitations and Future Work}

A limitation compared to other works is our need to train the ensemble model on feature vectors extracted from CSAI. Although it is a very light model, it is the only part of the system that cannot be distributed. Future work in this field could develop an ensemble model trained without CSAI~\cite{caetano2025neglected}; however, it would need to consider the risks of making a CSAI classification system publicly available to avoid misuse, such as searching for CSAI instead of eradicating~it.

This work is also limited by the use of Proxy Tasks created without CSAI classification in mind. Both models that were built in this context (Age/Gender and Few-Shot Scene Classification) had high values of feature importance, and future work in this area could explore training models for other Proxy Tasks, with a pipeline similar to the one proposed by \citet{Coelho2025MinimizingRisks}, and compare them to off-the-shelf state-of-the-art models for their respective tasks.

Moreover, we need to address skin tone as a target of interest. Firstly, the approach we followed in previous work yields a single scalar value, which does not allow the production of scatter plots for analysis. It also presented low feature importance on the aggregated measures. Skin tone or related targets, such as ethnicity, are relevant for producing statistics on apprehended CSAI, but within our approach, they do not contribute to the development of an explainable CSAI classification tool. 
Since ethnicity and skin color are protected attributes, specialized research, with ethical constraints in mind, is required to assess how to include these attributes in forensic contexts and whether their inclusion is justifiable.

Additionally, one unexplored area in this work is the evaluation of other approaches beyond gradient boosting to combine Proxy Tasks, such as statistical-based models or neural networks ensemble. Furthermore, the selected models for demographics-related Proxy Tasks (Age/Gender and ITA Skin Tone) cannot analyze images without visible faces. According to the authors of RCPD, $10\%$ of images in the dataset lack visible faces, and as a result, our pipeline is unable to extract demographic information from these images.

Furthermore, the Age/Gender model, although constructed in the context of CSAI, could be replaced by individual models for Age Estimation and Perceived Gender Prediction, allowing these two tasks to be analyzed separately for more detail in explainability. This could generate insight on whether binary perceived gender prediction truly is an important task, as it does not truly reflect the complex social construct that gender is, it does not generalize to different cultures, and is known to often have a lower accuracy for children.

Finally, although the system achieved competitive results on CSAI classification, a more descriptive analysis on how it could be directly used by LEA for both investigations and to comply with AI regulations depends on a participative design of a user interface, which is also an interesting future work.
\section{Conclusion}

In this work, we build upon and address limitations of the EISP framework. We applied it to real CSAI data with a novel selection of Proxy Tasks, achieving $91.9\%$ balanced accuracy on the validation set with the best-performing model combination, thereby achieving competitive performance in CSAI classification, while also providing the needed explainability to LEA through feature importance and visualization techniques. Both the application of PCA and the evaluation of Proxy Tasks combinations significantly optimized the feature vector storage and can be used to reduce storage usage and ensemble training and inference, if required. For all these reasons, using an ensemble of Proxy Tasks is a viable approach to building ethical and explainable CSAI classification systems.

\section*{Generative AI Usage Statement}
Generative AI was not used while writing this article. Writing assistance is limited to using the Tables Generator tool\footnote{\url{https://www.tablesgenerator.com}}, the native spelling correction on the Overleaf\footnote{\url{https://overleaf.com}} latex editing tool, and Google Translate for select sentences.

\section*{Author Contributions}
\textbf{Clara Ernesto}: Conceptualization, Methodology, Software, Visualization, Validation, Writing -- Original Draft; \textbf{Carlos Caetano}: Writing -- Review\& Editing; \textbf{Sandra Avila}: Project Administration, Funding Acquisition, Writing -- Review \& Editing, Supervision; \textbf{João Macedo}: Software, Investigation; \textbf{Camila Laranjeira}: Conceptualization, Methodology, Writing -- Original Draft, Supervision; \textbf{Leo S.~F.~Ribeiro}: Conceptualization, Supervision, Methodology, Resources, Writing -- Review \& Editing, Funding Acquisition.
\section*{Ethical Considerations Statement}
CSAI is a deeply harmful, sensitive and restricted domain, therefore no author who is not part of law enforcement agencies had access to the material for both mental health and legal reasons. 

We also thoroughly considered the environmental footprint while developing this work. The carbon emission and energy consumption of all experiments are reported in Section \ref{sec:env_footprint}, and the selection of lightweight models are not only a LEA restriction, but also a climatic requirement. As explained in Section \ref{sec:dino_explanation}, although our work experiments with computationally expensive image representation features, we favored DINO-vits8 instead of its newer versions, in order to reduce environmental impact, both inherited from its training and when applied to our pipeline. 

Finally, making public the information on how to create a CSAI classification system also creates risks, as a criminal maleficent reader who possess CSAI can try to reproduce related research and use this type of systems to search for CSAI in the internet, instead of eradicating it. This work mitigates this issue by not distributing models trained on sensitive data and lessening this risk through the use of non sensitive Proxy Tasks models.

\begin{acks}
This work is partially funded by FAPESP 2023/12086-9, and the Serrapilheira Institute R-2011-37776. C.~Ernesto is also funded by FAPESP 2025/08423-5. S.~Avila is also funded by FAPESP 2023/12865-8, 2020/09838-0, 2013/08293-7, H.IAAC 01245.003479/2024-10, and CNPq 316489/2023-9. 
\end{acks}

\bibliographystyle{ACM-Reference-Format}
\bibliography{arxiv}

@String{Computing = "Computing" }

@String{Computer = "{IEEE} Computer" }

@InProceedings{Caetano:SIBGRAPI:2024,
  author={Caetano, Carlos and Ferraz Ribeiro, Leo Sampaio and Laranjeira, Camila and dos Santos, Gabriel Oliveira and Barros, Artur and Petrucci, Caio and dos Santos, Andreza Aparecida and Macedo, Jo{\~a}o and Carvalho, Gil and Benevenuto, Fabricio and dos Santos, Jefersson A. and Avila, Sandra},
  booktitle={Conference on Graphics, Patterns and Images (SIBGRAPI)}, 
  title={Mastering Scene Understanding: Scene Graphs to the Rescue}, 
  year={2024},}

@ArtifactSoftware{R,
    title = {R: A Language and Environment for Statistical Computing},
    author = {{R Core Team}},
    organization = {R Foundation for Statistical Computing},
    address = {Vienna, Austria},
    year = {2019},
    url = {https://www.R-project.org/},
}

@article{zhang2020secretrevealergenerativemodelinversion,
  title={The Secret Revealer: Generative Model-Inversion Attacks Against Deep Neural Networks},
  author={Yuheng Zhang and R. Jia and Hengzhi Pei and Wenxiao Wang and Bo Li and Dawn Xiaodong Song},
  journal={IEEE/CVF Conference on Computer Vision and Pattern Recognition (CVPR)},
  year={2019},
}

@inproceedings{Coelho2025MinimizingRisks,
author = {Coelho, Thamiris and Ribeiro, Leo Sampaio Ferraz and Macedo, Jo\~{a}o and dos Santos, Jefersson A. and Avila, Sandra},
title = {Minimizing Risk Through Minimizing Model-Data Interaction: A Protocol For Relying on Proxy Tasks When Designing Child Sexual Abuse Imagery Detection Models},
year = {2025},
booktitle = {ACM Conference on Fairness, Accountability, and Transparency (FAccT)},
pages = {1543–1553},
}

@inproceedings{lundberg2017unifiedapproachinterpretingmodel,
author = {Lundberg, Scott M. and Lee, Su-In},
title = {A unified approach to interpreting model predictions},
year = {2017},
booktitle = {International Conference on Neural Information Processing Systems},
}

@inproceedings{Chen_2016,
   title={XGBoost: A Scalable Tree Boosting System},
   booktitle={ACM SIGKDD International Conference on Knowledge Discovery and Data Mining},
   author={Chen, Tianqi and Guestrin, Carlos},
   year={2016},
   pages={785–794},
   }

@article{zhou2017places,
   title={Places: A 10 million Image Database for Scene Recognition},
   author={Zhou, Bolei and Lapedriza, Agata and Khosla, Aditya and Oliva, Aude and Torralba, Antonio},
   journal={IEEE Transactions on Pattern Analysis and Machine Intelligence},
   year={2017},
   publisher={IEEE}
 }

@misc{AdamCodd/vit-base-nsfw-detector,
  title = {Vit Base NSFW Detector},
author = {AdamCodd},
  howpublished = {\url{https://huggingface.co/AdamCodd/vit-base-nsfw-detector}},
  note = {Accessed: 2025-08-10},
  year = {2025}
}

@software{Jocher_Ultralytics_YOLO_2023,
    author = {Jocher, Glenn and Qiu, Jing and Chaurasia, Ayush},
    license = {AGPL-3.0},
    month = {jan},
    title = {{Ultralytics YOLO}},
    url = {https://github.com/ultralytics/ultralytics},
    version = {8.0.0},
    year = {2023}
}

@misc{merler2019diversityfaces,
      title={Diversity in Faces}, 
      author={Michele Merler and Nalini Ratha and Rogerio S. Feris and John R. Smith},
      year={2019},
      eprint={1901.10436},
      archivePrefix={arXiv}, 
}

@inproceedings{laranjeira2022seeinglookinganalysispipeline,
author = {Laranjeira, Camila and Macedo, Jo{\~a}o and Avila, Sandra and dos Santos, Jefersson},
title = {Seeing without Looking: Analysis Pipeline for Child Sexual Abuse Datasets},
year = {2022},
booktitle = {ACM Conference on Fairness, Accountability, and Transparency (FAccT)},
}

@inproceedings{10.1007/978-3-030-59725-2_31,
    author = {Kinyanjui, Newton M. and Odonga, Timothy and Cintas, Celia and Codella, Noel C. F. and Panda, Rameswar and Sattigeri, Prasanna and Varshney, Kush R.},
    title = {Fairness of Classifiers Across Skin Tones in Dermatology},
    year = {2020},
    booktitle = {Medical Image Computing and Computer Assisted Intervention},
    pages = {320–329},
    }

@inproceedings{Macedo2018BenchmarkMethodologyChild,
  title = {A {{Benchmark Methodology}} for {{Child Pornography Detection}}},
  booktitle = {Conference on Graphics, Patterns and Images (SIBGRAPI)},
  author = {Macedo, Jo{\~a}o and Costa, Filipe and {A. dos Santos}, Jefersson},
  year = {2018},
  pages = {455--462},
  publisher = {IEEE},
}

@software{ivan_de_paz_centeno_2024_13901379,
  author       = {Iván de Paz Centeno},
  title        = {ipazc/mtcnn: v1.0.0},
  month        = oct,
  year         = 2024,
  publisher    = {Zenodo},
  version      = {v1.0.0},
  doi          = {10.5281/zenodo.13901379},
  url          = {https://doi.org/10.5281/zenodo.13901379},
}

@inproceedings{10.1145/3292522.3326027,
author = {Reis, Julio C. S. and Correia, Andr\'{e} and Murai, Fabr\'{\i}cio and Veloso, Adriano and Benevenuto, Fabr\'{\i}cio},
title = {Explainable Machine Learning for Fake News Detection},
year = {2019},
booktitle = {ACM Conference on Web Science},
pages = {17–26},
}

@article{mcinnes2020umapuniformmanifoldapproximation,
year = {2018},
author = {McInnes, Leland and Healy, John and Saul, Nathaniel and Großberger, Lukas},
title = {UMAP: Uniform Manifold Approximation and Projection},
journal = {Journal of Open Source Software}
}

@inproceedings{lottick2019energyusagereportsenvironmental,
  author    = {Lottick, Kadan and Susai, Silvia and Friedler, Sorelle A. and Wilson, Jonathan P.},
  title     = {Energy Usage Reports: Environmental awareness as part of algorithmic accountability},
  booktitle = {NeurIPS Workshop on Tackling Climate Change with Machine Learning},
  year      = {2019},
}

@article{valois2025leveraging,
  title={Leveraging self-supervised learning for scene classification in child sexual abuse imagery},
  author={Valois, Pedro H. V. and Macedo, Jo{\~a}o and Ribeiro, Leo Sampaio Ferraz and dos Santos, Jefersson A and Avila, Sandra},
  journal={Forensic Science International: Digital Investigation},
  volume={53},
  pages={301918},
  year={2025},
  publisher={Elsevier}
}

@article{macedo2025child,
  title={Child Sexual Abuse Datasets: A Systematic Review},
  author={Macedo, Jo{\~a}o and Laranjeira, Camila and Ribeiro, Leo Sampaio Ferraz and Caetano, Carlos and Benevenuto, Fabricio and Avila, Sandra and dos Santos, Jefersson A},
  journal = {Research Square},
  doi={https://doi.org/10.21203/rs.3.rs-7963252/v1},
  year={2025}
}

@InProceedings{DeCastro2010NudetectiveCSAM,
  author    = {{Polastro}, Mateus and {Eleuterio}, Pedro},
  booktitle = {Workshops on Database and Expert Systems Applications},
  title     = {Nudetective: {{A}} Forensic Tool to Help Combat Child Pornography through Automatic Nudity Detection},
  year      = {2010},
}

@InProceedings{DeCastro2012StatisticalCSAM,
  author    = {{Polastro}, Mateus and {Eleuterio}, Pedro},
  booktitle = {International Conference on Availability, Reliability and Security},
  title     = {A Statistical Approach for Identifying Videos of Child Pornography at Crime Scenes},
  year      = {2012},
}

@article{Gangwar2021AttMCNNAttentionMetric,
  title = {{{AttM-CNN}}: {{Attention}} and Metric Learning Based {{CNN}} for Pornography, Age and {{Child Sexual Abuse}} ({{CSA}}) {{Detection}} in Images},
  shorttitle = {{{AttM-CNN}}},
  author = {Gangwar, Abhishek and {Gonz{\'a}lez-Castro}, V{\'i}ctor and Alegre, Enrique and Fidalgo, Eduardo},
  year = {2021},
  journal = {Neurocomputing},
  volume = {445},
  pages = {81--104},
}

@article{Maaten08tsne,
  title = {Visualizing Data Using T-{{SNE}}},
  author = {{van der Maaten}, Laurens and Hinton, Geoffrey},
  year = {2008},
  journal = {Journal of Machine Learning Research},
  volume = {9},
  number = {86},
  pages = {2579--2605}
}

@misc{Oquab2023DINOv2LearningRobust,
  title = {{{DINOv2}}: {{Learning Robust Visual Features}} without {{Supervision}}},
  shorttitle = {{{DINOv2}}},
  author = {Oquab, Maxime and Darcet, Timoth{\'e}e and Moutakanni, Th{\'e}o and Vo, Huy and Szafraniec, Marc and Khalidov, Vasil and Fernandez, Pierre and Haziza, Daniel and Massa, Francisco and {El-Nouby}, Alaaeldin and Assran, Mahmoud and others},
  year = {2023},
  number = {arXiv:2304.07193},
  eprint = {2304.07193},
  publisher = {{arXiv}},
}

@misc{PhotoDNA,
  title = {Tackling Child Sexual Abuse Strategy.},
  author = {{Microsoft}},
  year = {2009}
}

@InProceedings{SaeBae2014TowardsAutomaticCSAM,
  author    = {{Sae-Bae}, Napa and Sun, Xiaoxi and Sencar, Husrev T and Memon, Nasir D},
  booktitle = {{{IEEE}} International Conference on Image Processing},
  title     = {Towards Automatic Detection of Child Pornography},
  year      = {2014},
}

@article{Vitorino2018LeveragingCSAM,
  title = {Leveraging Deep Neural Networks to Fight Child Pornography in the Age of Social Media},
  author = {Vitorino, Paulo and Avila, Sandra and Perez, Mauricio and Rocha, Anderson},
  year = {2018},
  journal = {Journal of Visual Communication and Image Representation},
  volume = {50},
  pages = {303--313},
}

@inproceedings{Westlake2012ComparingCSAM,
  title = {Comparing Methods for Detecting Child Exploitation Content Online},
  booktitle = {European Intelligence and Security Informatics Conference},
  author = {Westlake, Bryce and Bouchard, Martin and Frank, Richard},
  year = {2012},
}

@inproceedings{Yiallourou2017DetectionImagesContaining,
  title = {On the Detection of Images Containing Child-Pornographic Material},
  booktitle = {{{International Conference}} on {{Telecommunications}}},
  author = {Yiallourou, Emilios and Demetriou, Rafaella and Lanitis, Andreas},
  year = {2017},
  pages = {1--5},
  publisher = {{IEEE}},
}

@InProceedings{Struppek2022LearningBreakDeep,
  author    = {Struppek, Lukas and Hintersdorf, Dominik and Neider, Daniel and Kersting, Kristian},
  booktitle = {{{ACM Conference}} on {{Fairness}}, {{Accountability}}, and {{Transparency}}},
  title     = {Learning to {{Break Deep Perceptual Hashing}}: {{The Use Case NeuralHash}}},
  year      = {2022},
}

@inproceedings{caetano2025neglected,
  title={Neglected Risks: The Disturbing Reality of Children's Images in Datasets and the Urgent Call for Accountability},
  author={Caetano, Carlos and Santos, Gabriel O dos and Petrucci, Caio and Barros, Artur and Laranjeira, Camila and Ribeiro, Leo Sampaio Ferraz and de Mendon{\c{c}}a, J{\'u}lia F and Santos, Jefersson A dos and Avila, Sandra},
  booktitle={ACM Conference on Fairness, Accountability, and Transparency (FAccT)},
  year={2025}
}

@inproceedings{coelho2024transformers,
  title={Transformers-Based Few-Shot Learning for Scene Classification in Child Sexual Abuse Imagery},
  author={Coelho, Thamiris and Ribeiro, Leo Sampaio Ferraz and Macedo, Jo{\~a}o and dos Santos, Jefersson A and Avila, Sandra},
  booktitle={Conference on Graphics, Patterns and Images (SIBGRAPI)},
  pages={8--14},
  year={2024},
  organization={SBC}
}

@inproceedings{EISPErnesto2025,
  title={Proxy Tasks Ensemble for Explainable Inference in Sensitive Data},
  author={Ernesto, Clara and Avila, Sandra and Caetano, Carlos and Ribeiro, Leo Sampaio Ferraz},
  booktitle={Conference on Graphics, Patterns and Images (SIBGRAPI)},
  pages={194--199},
  year={2025},
  organization={SBC}
}

@inproceedings{optuna_2019,
	title={Optuna: A Next-generation Hyperparameter Optimization Framework},
	author={Akiba, Takuya and Sano, Shotaro and Yanase, Toshihiko and Ohta, Takeru and Koyama, Masanori},
	booktitle={{ACM} {SIGKDD} International Conference on Knowledge Discovery and Data Mining},
	year={2019}
}

@inproceedings{caron2021emergingpropertiesselfsupervisedvision,
  title={Emerging Properties in Self-Supervised Vision Transformers},
  author={Mathilde Caron and Hugo Touvron and Ishan Misra and Herv\'e J\'egou and Julien Mairal and Piotr Bojanowski and Armand Joulin},
  booktitle={IEEE/CVF International Conference on Computer Vision},
  year={2021},
}

@article{sanchez2019practitioner,
  title={A practitioner survey exploring the value of forensic tools, AI, filtering, \& safer presentation for investigating child sexual abuse material (CSAM)},
  author={Sanchez, Laura and Grajeda, Cinthya and Baggili, Ibrahim and Hall, Cory},
  journal={Digital Investigation},
  volume={29},
  pages={S124--S142},
  year={2019},
  publisher={Elsevier}
}

@inproceedings{Walke2023AIAct,
author = {Walke, Acting Professor Dr. Fabian and Bennek, Lars and Winkler, Till},
year = {2023},
month = {09},
pages = {},
title = {Artificial Intelligence Explainability Requirements of the AI Act and Metrics for Measuring Compliance}
}

@inproceedings{barros2025attention,
  author    = {Barros, Artur and Caetano, Carlos and Macedo, Jo{\~a}o and Santos, Jefersson A. dos and Avila, Sandra},
  title     = {Attention over Scene Graphs: Indoor Scene Representations Toward CSAI Classification},
  booktitle = {Proceedings of the British Machine Vision Conference (BMVC) Workshops},
  year      = {2025}
}

@article{ORONOWICZJASKOWIAK2024102619,
title = {Using expert-reviewed CSAM to train CNNs and its anthropological analysis},
journal = {Journal of Forensic and Legal Medicine},
volume = {101},
pages = {102619},
year = {2024},
author = {Wojciech Oronowicz-Jaśkowiak and Tomasz Kozłowski and Marta Polańska and Jerzy Wojciechowski and Piotr Wasilewski and Dominik Ślęzak and Mirosław Kowaluk}
}

@article{CastrillnSantana2017EvaluationOL,
  title={Evaluation of local descriptors and CNNs for non-adult detection in visual content},
  author={Modesto Castrill{\'o}n-Santana and Javier Lorenzo-Navarro and Carlos Manuel Travieso-Gonz{\'a}lez and David Freire-Obreg{\'o}n and Jes{\'u}s B. Alonso},
  journal={Pattern Recognit. Lett.},
  year={2017},
  volume={113},
  pages={10-18},
}

@article{HORNOR2010358,
title = {Child Sexual Abuse: Consequences and Implications},
journal = {Journal of Pediatric Health Care},
volume = {24},
number = {6},
pages = {358-364},
year = {2010},
issn = {0891-5245},
author = {Gail Hornor}
}

@inproceedings{schulze2014automatic,
  title={Automatic detection of CSA media by multi-modal feature fusion for law enforcement support},
  author={Schulze, Christian and Henter, Dominik and Borth, Damian and Dengel, Andreas},
  booktitle={International Conference on Multimedia Retrieval},
  pages={353--360},
  year={2014}
}

@article{kloess2019challenges,
  title={The challenges of identifying and classifying child sexual abuse material},
  author={Kloess, Juliane A and Woodhams, Jessica and Whittle, Helen and Grant, Tim and Hamilton-Giachritsis, Catherine E},
  journal={Sexual Abuse},
  volume={31},
  number={2},
  pages={173--196},
  year={2019},
  publisher={Sage Publications Sage CA: Los Angeles, CA}
}

@misc{breder_2025_17247278,
  author       = {Breder, Gabriel and
                  Ferro, Mariza},
  title        = {wAIter: Serving Awareness - Environmental
                   Footprint Estimator for Artificial Intelligence
                   Models
                  },
  month        = oct,
  year         = 2025,
  publisher    = {Zenodo},
  doi          = {10.5281/zenodo.17247278},
  url          = {https://doi.org/10.5281/zenodo.17247278},
}

@misc{simeoni2025dinov3,
      title={DINOv3}, 
      author={Oriane Siméoni and Huy V. Vo and Maximilian Seitzer and Federico Baldassarre and Maxime Oquab and Cijo Jose and Vasil Khalidov and Marc Szafraniec and Seungeun Yi and Michaël Ramamonjisoa and Francisco Massa and Daniel Haziza and Luca Wehrstedt and others},
      year={2025},
      eprint={2508.10104},
      archivePrefix={arXiv},
      url={https://arxiv.org/abs/2508.10104}, 
}

@article{unicef2024numbers,
  title={When Numbers Demand Action: Confronting the global scale of sexual violence against children},
  author={UNICEF and others},
  journal={New York},
  year={2024},
}

\appendix{
\section{Holdout Experiment Feature Importance Plots}
\label{sec:crossval_featimp}
\begin{figure}[h]
    \centering
    \includegraphics[width=\linewidth]{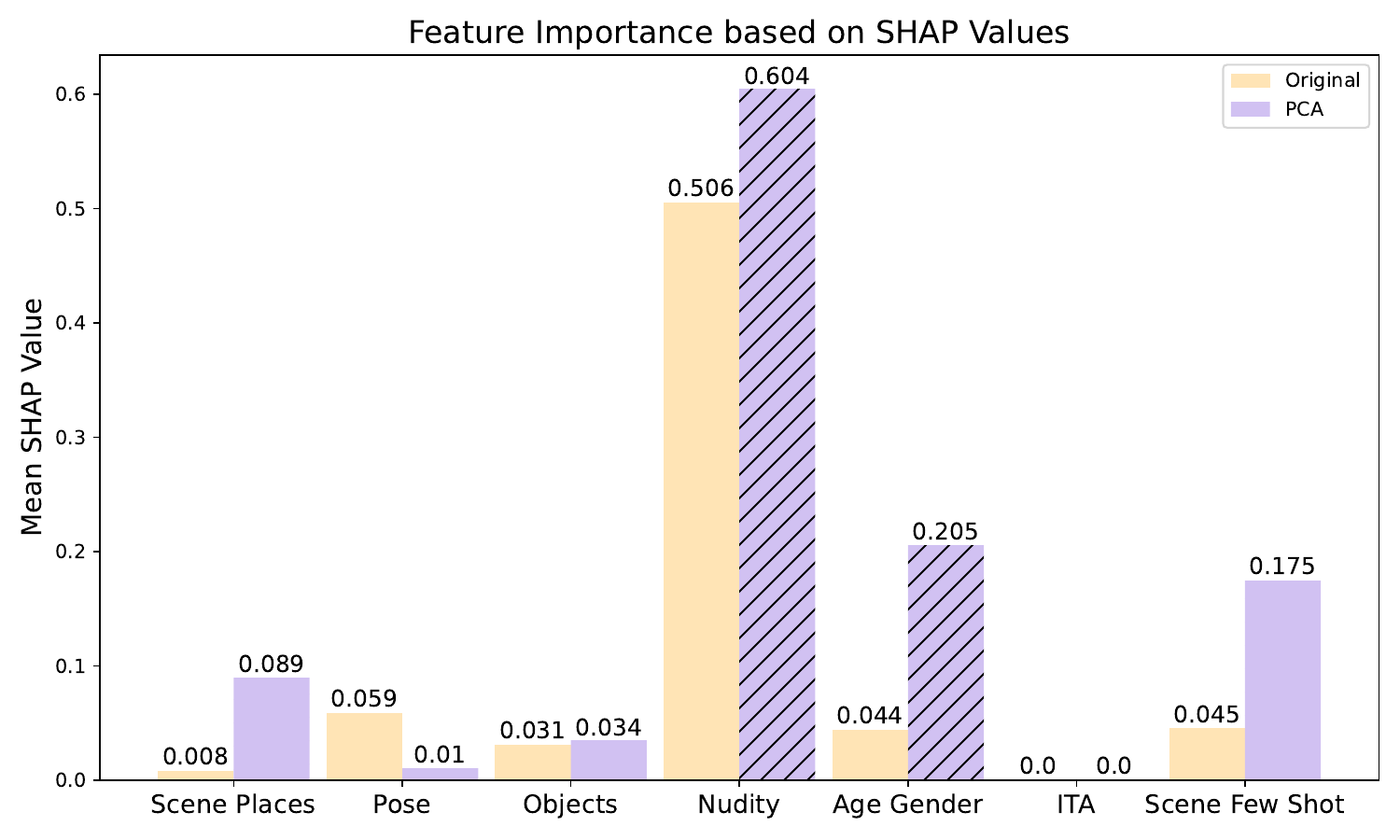}
     \caption{Average feature importance (aggregated SHAP values) for the Holdout experiment, using both PCA-applied and Original features.}
     \Description{Average feature importance (aggregated SHAP values) for the Holdout experiment, using both PCA-applied and Original features. The models which show greater feature importance are the Nudity, Age Gender and Scene Few Shot model}
    \label{fig:simple_featimp}
\end{figure}

\section{UMAP Plot Per Proxy Task Model}
\label{sec:umap_per_model}

\begin{figure*}[h]
     \centering
     \begin{subfigure}[b]{0.4\textwidth}
         \centering
         \includegraphics[width=\linewidth,trim={2.6cm 2cm 2.1cm 2cm},clip]{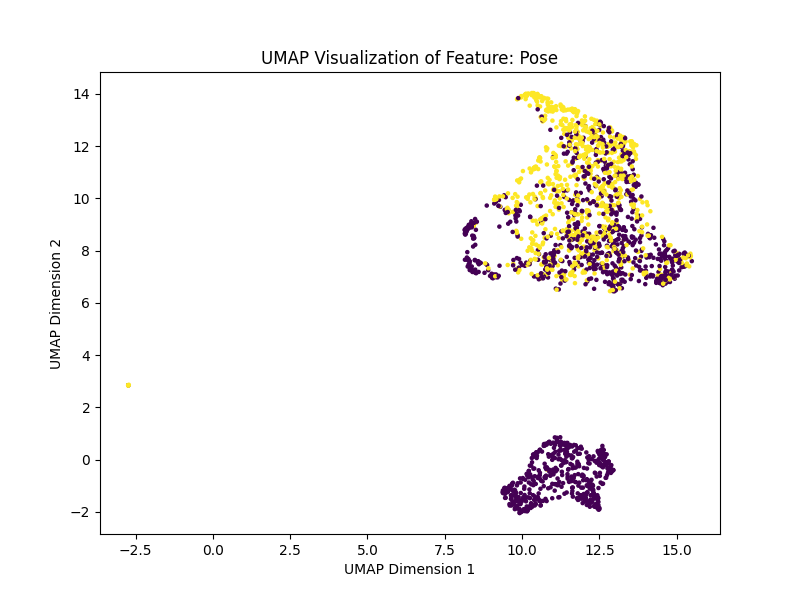}
        \caption{Pose detection}
        \label{fig:umap_pose}
     \end{subfigure}
     \hfill
     \begin{subfigure}[b]{0.4\textwidth}
         \centering
          \includegraphics[width=\linewidth,trim={2.6cm 2cm 2.1cm 2cm},clip]{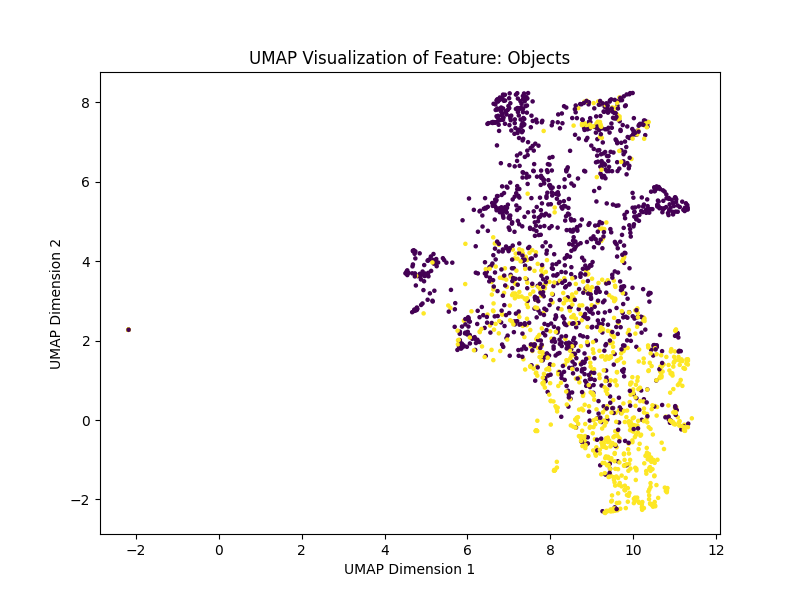}
        \caption{Object detection}
        \label{fig:umap_object}
     \end{subfigure}
     \hfill
     \begin{subfigure}[b]{0.4\textwidth}
         \centering
         \includegraphics[width=\linewidth,trim={2.6cm 2cm 2.1cm 2cm},clip]{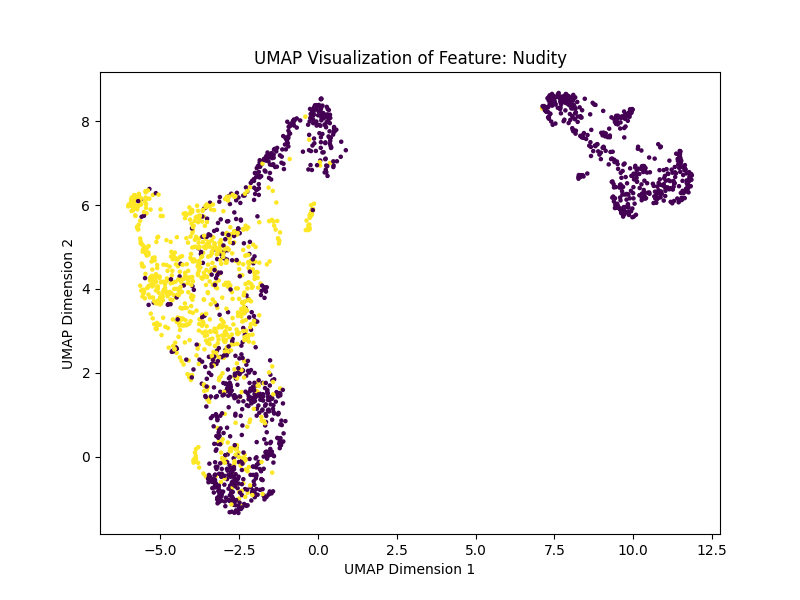}    
         \caption{Nudity classification}
         \label{fig:umap_nudity}
     \end{subfigure}
     \hfill
     \begin{subfigure}[b]{0.4\textwidth}
         \centering
         \includegraphics[width=\linewidth,trim={2.6cm 2cm 2.1cm 2cm},clip]{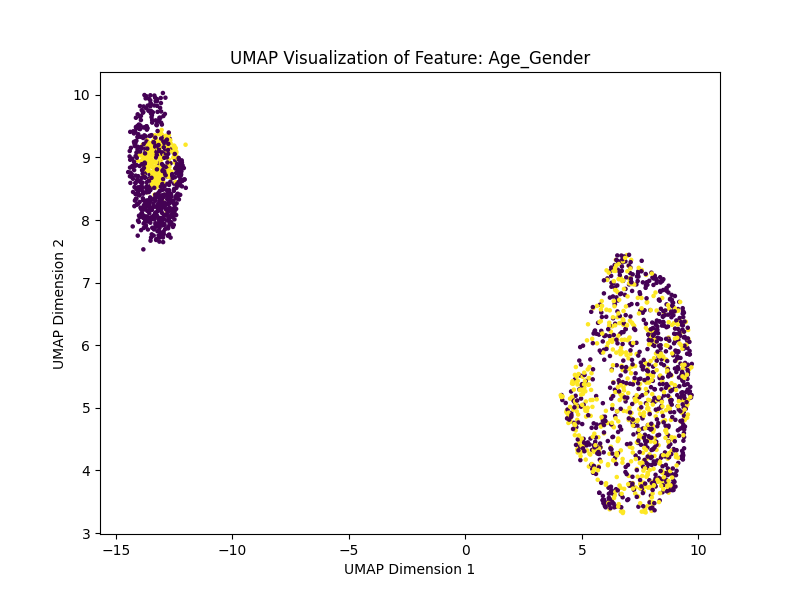}
        \caption{Age and perceived gender recognition}
        \label{fig:umap_agegender}
     \end{subfigure}
     \hfill
     \begin{subfigure}[b]{0.4\textwidth}
         \centering
         \includegraphics[width=\linewidth,trim={2.6cm 2cm 2.1cm 2cm},clip]{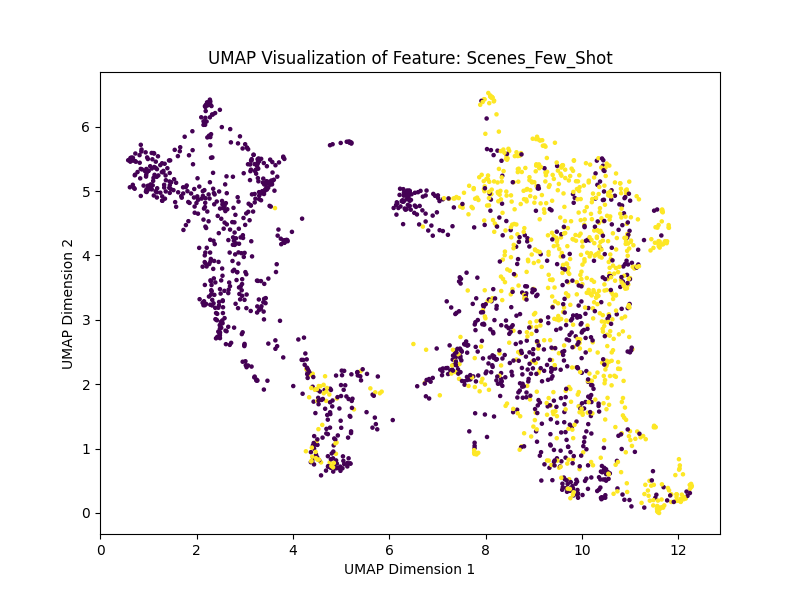}
        \caption{Few-shot scene classification}
        \label{fig:umap_fewshot}
     \end{subfigure}
     \hfill
     \begin{subfigure}[b]{0.4\textwidth}
         \centering
          \includegraphics[width=\linewidth,trim={2.6cm 2cm 2.1cm 2cm},clip]{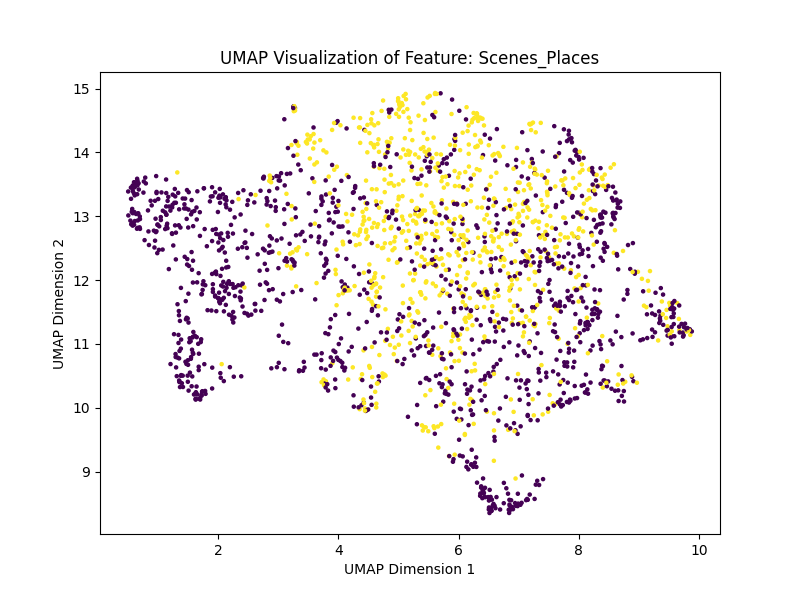}    
          \caption{Places scene classification}
          \label{fig:umap_places}
     \end{subfigure}
     \caption{UMAP projections of the feature space for each Proxy Task model after PCA preprocessing. Yellow dots represent CSAI images and purple dots represent Non-CSAI images.}
     \Description{UMAP projections of the feature space for each Proxy Task model after PCA preprocessing. Yellow dots represent CSAI images and purple dots represent Non-CSAI images. Most models show a clear division between yellow and purple dots, except the Age and perceived gender model.}
    \label{fig:umap_features}
\end{figure*}


\clearpage

\end{document}